\documentclass[aps,prx,twocolumn,10pt,amsmath,amssymb,floatfix,superscriptaddress]{revtex4-2}

\usepackage{graphicx} 
\usepackage[caption=false]{subfig}

\usepackage{bm} 
\usepackage{braket,bbold}
\usepackage{hyperref}
\usepackage{multirow}
\usepackage[table]{xcolor}
\newcommand{\bs}[1]{\boldsymbol{#1}}

\begin{document}

\title{Topological Transitions of Evanescent Waves}

\author{Ignacio Pavez}
\affiliation{Blackett Laboratory, Imperial College London, London SW7 2AZ, United Kingdom}

\author{Tom Sheppard}
\affiliation{School of Physics and Astronomy, University of Birmingham, Birmingham, United Kingdom}

\author{Yoonseok Hwang}
\affiliation{Blackett Laboratory, Imperial College London, London SW7 2AZ, United Kingdom}

\author{Hannah M. Price}
\affiliation{School of Physics and Astronomy, University of Birmingham, Birmingham, United Kingdom}

\author{Frank Schindler}
\affiliation{Blackett Laboratory, Imperial College London, London SW7 2AZ, United Kingdom}

\begin{abstract}
We study the band structure of evanescent waves, that is, the decaying or growing solutions that live within the gaps of tight-binding Hamiltonians. Although the underlying Hamiltonian is Hermitian, these modes carry complex wavevectors and therefore exhibit a non-Hermitian dispersion. Using the transfer matrix formalism in one dimension, we show that, in addition to the standard exceptional points that appear at the band edges, a different type of exceptional point can arise inside the gap, protected by spinless time-reversal symmetry. Tuning model parameters drives a topological transition where such in-gap exceptional points are nucleated in pairs, mediated by a crossing between purely imaginary momentum branches. We show that this spectral structure is diagnosed by a $\mathbb{Z}_2$ index assigned to every energy inside the gap, with the in-gap exceptional points arising as its domain walls. A generic bound state in the system has a tail that is purely exponential when its energy lies between a band edge and the nearest exceptional point, but becomes oscillatory -- with a wavevector differing from the band-edge momentum -- when its energy lies between the pair of in-gap exceptional points. We illustrate these results using continuum and tight-binding models, as well as an anisotropic photonic crystal.
\end{abstract}

\maketitle
\section{Introduction}
Band topology has become a central organising principle in condensed matter physics: global properties of Bloch eigenstates can determine robust physical phenomena, from quantised Hall conductance to protected edge states in topological insulators and topological superconductors~\cite{vonKlitzing1980,Laughlin1981,NiuThoulessWu1985,TKNN1982,Kohmoto1985,KaneMele2005_Z2,BernevigHughesZhang2006,FuKaneMele2007,HasanKane2010,QiZhang2011,SatoAndo2017}. In crystalline systems, this result is usually formulated in terms of the Bloch Hamiltonian $H_{\mathrm{B}}(k)$, whose eigenstates over the Brillouin zone give rise to topological invariants and the bulk--boundary correspondence~\cite{Bloch1929,Berry1984,TKNN1982,Hatsugai1993PRL,XiaoChangNiu2010,HasanKane2010}.
However, the Bloch Hamiltonian description primarily focuses on the properties of bulk bands and provides limited direct information about the structure of spectral gaps themselves~\cite{Kohn1959,ChangSchulman1982}. In particular, states with energies inside a bulk gap, such as those arising at boundaries, defects, or from local perturbations, correspond to spatially decaying wavefunctions and therefore cannot be described by a real crystal momentum~\cite{Shockley1939,Heine1964,ChangSchulman1982}. From the Hamiltonian perspective, the spatial profile of such states, far away from the boundary or defect, can be obtained by solving
\begin{equation}
\label{eq: det equation for H}
    \det \left[ H_{\mathrm{B}}(k)-E \right]=0,
\end{equation}
where $E$ is the energy of the state~\cite{ChangSchulman1982}. While this equation yields real momenta $k$ for energies within the bulk bands, for energies inside a spectral gap the solutions involve complex momenta corresponding to evanescent modes~\cite{Kohn1959,ChangSchulman1982}. Determining these complex solutions is generally nontrivial: even for a multiband Hamiltonian, obtaining the complete set of complex momenta at fixed energy amounts to finding all roots of a nonlinear equation. Moreover, approximations based on a local expansion around a band edge do not capture the global structure of the complex-momentum spectrum inside the gap.

A framework for addressing these questions in one dimension is provided by the transfer matrix formalism, which describes the evolution of wavefunctions along a given direction, whether in space or in time~\cite{DwivediChua2016,asgari2024photonic}. Regarding the spatial case, instead of solving implicitly for complex momenta, the transfer matrix reduces the problem to an eigenvalue equation whose spectrum determines the Bloch momentum given a fixed energy. In this way, both bulk plane waves and bound-state tails can be obtained from a diagonalisation of the transfer matrix~\cite{DwivediChua2016,Hatsugai1993PRB}. This approach has been used to derive edge states analytically in general non-interacting tight-binding systems and provides a direct connection between edge states and bulk topological invariants~\cite{DwivediChua2016,Hatsugai1993PRL}.

The analysis of states inside spectral gaps naturally leads to non-Hermitian physics~\cite{Kohn1959,BergholtzBudichKunst2021}. When energies lie within a gap, the corresponding complex momenta imply that the Bloch Hamiltonian evaluated at these momenta becomes non-Hermitian~\cite{KunstDwivedi2019,YaoWang2018}. The spectral structure associated with these complex solutions can therefore exhibit characteristic non-Hermitian features such as robust exceptional points, where eigenvalues and eigenvectors coalesce~\cite{Kato1966,Heiss2012,BergholtzBudichKunst2021,saha2023universal,saha2023exceptional,saha2025arbitrary}. Such structures encode information about the analytic structure of the gap and the properties of localised states~\cite{Hatsugai1993PRB,Kohn1959}.

In this work we focus on gapped band structures (insulators) with spinless time-reversal symmetry (TRS). In the transfer-matrix formulation, spinless TRS can be represented as a reality condition, while Hermiticity of the underlying Hamiltonian imposes an additional complex symplectic constraint. The corresponding class of real symplectic operators has previously been studied within Krein-space theory~\cite{schulz2017signatures}. More broadly, spinless TRS plays a central role in condensed matter physics, where it corresponds to the absence of a magnetic field and spin-orbit coupling. More recently, this symmetry entered the classification of non-Hermitian phases and exceptional degeneracies, where it can protect or constrain robust exceptional points in complex spectra ~\cite{Gong2018,Kawabata2019PRX,Kawabata2019PRL, BergholtzBudichKunst2021}.

The connection between localised states and complex momentum has a long history in the theory of surface and edge states ~\cite{Shockley1939,Heine1964,Kohn1959,ChangSchulman1982, Hatsugai1993PRB,DwivediChua2016,reuter2017unified}. In these works, surface- or boundary-localised wavefunctions are described in terms of evanescent bulk solutions, equivalently complex Bloch factors or transfer-matrix eigenmodes. More recently, complex-Brillouin-zone methods have been used to show that the local oscillations of evanescent waves can change qualitatively depending on where inside a band gap the frequency lies~\cite{debruijn2025complex,de2026complex}.
Related work has also classified Fermi-arc surface states of Weyl semimetals according to whether their evanescent tails exhibit oscillatory or purely exponential decay, in terms of exceptional points of the analytically continued Hamiltonian at fixed energy~\cite{gonzalez2017topological}. What remains less clear is what, if anything, protects this structure, and how it is organised by the symmetries of the underlying Hermitian problem. Related links between bound states and exceptional points have been explored in genuinely non-Hermitian wave systems, especially in photonic platforms where bound states in the continuum can be tuned into or merged with exceptional points ~\cite{Kang2023BIC,CanosValero2025ExceptionalBIC, Wang2025PhotoswitchableEPBIC}. For Hermitian TRS systems, it therefore remains to be understood how TRS constrains the in-gap exceptional points of the transfer matrix, in what sense they are topologically protected, and whether they leave observable signatures in conventional bound states.

In this work, we investigate the general spectral structure of gaps in one-dimensional (1D) Hermitian systems with spinless TRS from the transfer-matrix perspective. We show that the robust exceptional points that occur inside a spectral gap are protected by the combination of spinless TRS and the symplectic condition implied by Hermiticity, and are diagnosed by a $\mathbb{Z}_2$ index. These exceptional points can be created or annihilated in pairs without closing the propagating bulk gap and delimit the energy region in which generic bound-state tails change from purely exponential decay to exponentially damped oscillations. By realising the same mechanism in an anisotropic photonic crystal, we further show that it is not restricted to tight-binding systems, but extends to periodic systems whose transfer matrices obey the same symmetry constraints. Since the decay of bound states is determined by the transfer matrix alone, the same transition between purely exponential and oscillatory decay is expected for localised defect modes in photonic crystals.

\subsection{Review of the transfer matrix}
\label{sec: transfer matrix}
In this section, we briefly review the construction of the transfer matrix for a non-interacting 1D tight-binding model, following Ref.~\cite{DwivediChua2016}, and introduce the notation used throughout this work. We begin by considering a bulk translationally invariant 1D lattice with hopping range $R$ and open boundary conditions, and a Hermitian Hamiltonian
\begin{equation}
H=\sum_{n=1}^{N}\sum_{\alpha,\beta=1}^{q}\sum_{l=0}^{R}t_{l,\alpha,\beta}\ket{n,\alpha}\bra{n+l,\beta}+\mathrm{h.c.},
\end{equation}
where $N$ is the number of lattice sites, each with $q$ internal degrees of freedom. For open boundary conditions, states with $n+l>N$ are omitted. In this basis, wavefunctions are written as
\begin{equation}
    \ket{\psi}=\sum_{n=1}^{N}\sum_{\alpha=1}^{q} \psi_{n,\alpha}\ket{n,\alpha}.
\end{equation}
Then, the time-independent Schrödinger equation $H \ket{\psi} = E \ket{\psi}$ becomes
\begin{equation}
\label{eq: time-ind SE}
    \sum_{l=0}^{R} \left(t_l^{\dagger}\psi_{n-l} + t_l \psi_{n+l}\right)=E\psi_n,
\end{equation}
where $t_l$ are the matrices whose components are $t_{l,\alpha,\beta}$ and $\psi_{n}$ is the vector in $\mathbb{C}^{q}$ whose components are $\psi_{n,\alpha}$. 
As an example, for the case $q=R=1$, we can solve for $\psi_{n+1}$ in Eq.~\eqref{eq: time-ind SE} and thus construct the equation
\begin{equation}
\label{eq: transf matrix eq for non-sing J}
\begin{aligned}
    \begin{pmatrix}
        \psi_{n+1}\\ \psi_{n}
    \end{pmatrix}
    &=T(E)\begin{pmatrix}
        \psi_{n}\\ \psi_{n-1}
    \end{pmatrix},\quad     T(E)=\begin{pmatrix}
        \frac{E-\tilde{t}_0}{t_1}&-\frac{t_1^*}{t_1}\\
        1&0
    \end{pmatrix},\\
    \end{aligned}
\end{equation}
where $T(E)$ is the transfer matrix and $\tilde{t}_0:=t_0+t_0^{\dagger}$. 

Any 1D lattice with a finite number of orbitals $q$ and finite hopping range $R$ can be mapped onto an effective nearest-neighbour system by grouping sites into \emph{supercells}, as discussed in Ref.~\cite{DwivediChua2016}. In this representation, Eq.~\eqref{eq: time-ind SE} becomes
\begin{equation}
\label{eq: NNH for supercells}
    J^{\dagger}\Psi_{n-1}+M\Psi_{n}+J\Psi_{n+1}=E\Psi_{n},
\end{equation}
where $\Psi_{n}$ represents the wavefunction component at supercell site $n$, $J$ is the hopping matrix and $M$ is the onsite matrix. In particular, the smallest representation corresponds to constructing $\Psi_{n}$ as
\begin{equation} \label{psin def}
    \Psi_{n}:=\begin{pmatrix}
        \psi_{(n-1)R+1}\\
        \psi_{(n-1)R+2}\\
        \vdots\\
        \psi_{nR}
    \end{pmatrix}\in\mathbb{C}^{qR},
\end{equation}
so that $J$ becomes a block lower-triangular matrix
\begin{equation}
    J=\begin{pmatrix}
        t_R&0&0&0&\ldots&0 \\
        t_{R-1}&t_R&0&0&\ldots &0\\
        t_{R-2}&t_{R-1}&t_R&0&\ldots &0\\
        \vdots&\vdots&\vdots&\vdots&\ddots & \vdots\\
        t_1&t_2&t_3&t_4&\ldots&t_R
    \end{pmatrix}.
\end{equation}
We note that $J$ is singular if and only if $t_R$ is singular. This corresponds to a situation where there is a degree of freedom at position $n + R$ that is not connected with site $n$. 
Assuming $J$ is nonsingular, we now construct the vector
\begin{equation}
\label{eq: def of Phi for J nonsingular}
    \Phi_{n}:=\begin{pmatrix}
        \Psi_{n+1}\\ \Psi_{n}
    \end{pmatrix}
\end{equation}
to construct the transfer matrix via
\begin{equation}
    \Phi_{n+1} = T(E)\Phi_{n}.
\end{equation}
We will also use this equation to define the transfer matrix in a general setting where the specific form of $\Phi_{n}$ may differ. For example, when $J$ is singular, Eq.~\eqref{eq: def of Phi for J nonsingular} contains redundant information, since some of the degrees of freedom in $\Psi_{n}$ do not enter into the relation given by Eq.~\eqref{eq: NNH for supercells}. In this case, we follow Dwivedi and Chua~\cite{DwivediChua2016} in constructing the transfer matrix. Their prescription is reproduced in Appendix~\ref{appendix: construction general transfer matrix} for completeness. Furthermore, in the case of a continuum model, the corresponding vector $\Phi_{n}$ is simply the wavefunction evaluated at position $x$, $\Phi_{n}\rightarrow \psi(x)$.

For a translationally invariant system with unit period, the transfer matrix determines the allowed complex momenta $k$ as functions of $E$ through the eigenvalue equation
\begin{equation}
    T(E)\Phi_{n}=e^{\mathrm{i}k(E)}\Phi_{n}.
\end{equation}
We now investigate the spectral features of  $T(E)$.

\section{In-gap topological transitions}
\label{sec: in-gap transition}
We define the Bloch basis as
\begin{equation}
    \ket{k,\alpha}=\frac{1}{\sqrt{N}}\sum_{n=1}^{N}e^{\mathrm{i}kn}\ket{n,\alpha},
\end{equation}
where $k$ lies in the first Brillouin zone. In this basis the Hamiltonian decomposes as
\begin{equation}
    H=\sum_{k\in \mathrm{BZ}}\sum_{\alpha,\beta=1}^{q}
    \left[H_{\mathrm{B}}(k)\right]_{\alpha\beta}
    \ket{k,\alpha}\bra{k,\beta},
\end{equation}
where $H_{\mathrm{B}}(k)$ is a $q\times q$ Hermitian matrix acting on the internal degrees of freedom. The band structure is obtained by diagonalising $H_{\mathrm{B}}(k)$ for real $k$ in the first Brillouin zone, $k \in [0,2\pi)$ for unit lattice spacing.

To understand how a spectral gap can acquire internal structure, we first focus on the local mechanism by which a gap is created. The simplest situation is a linear crossing of bands that is gapped out by a perturbation. The physics sufficiently close to such a crossing is controlled by the local expansion of the Bloch Hamiltonian, and this provides a natural starting point for classifying possible gap openings.
We work in the thermodynamic limit, so that the crystal momentum can be treated as a continuous variable. Let $k_0$ be a momentum at which two or more bands cross before the gap is opened, and define $p=k-k_0$. Assuming local hoppings, the Bloch Hamiltonian is a smooth function of $k$ in the neighbourhood of $k_0$, so that we may Taylor expand it as
\begin{equation}
    H_{\mathrm{B}}(k_0+p)
    =
    H_{\mathrm{B}}(k_0)
    +
    p\,\frac{\mathrm{d}H_{\mathrm{B}}}{\mathrm{d}k}(k_0)
    +O(p^2).
\end{equation}
The variable $p$ is the crystal momentum relative to $k_0$: it is not a periodic Brillouin-zone coordinate, and the expansion is valid only for sufficiently small $|p|$. After shifting the energy of the crossing to zero and restricting to the subspace of bands involved in the crossing, the leading long-wavelength Hamiltonian has a Dirac form
\begin{equation}
    H_{\mathrm{D}}(p)=p\gamma+m,
\end{equation}
where $\gamma$ comes from the linear term in the Taylor expansion, while $m$ comes from the constant term and opens the gap. For simplicity, we restrict to Dirac models with invertible $\gamma$.

This local Dirac description is useful because the complex-momentum dispersion near a small gap is controlled by the same data that determine how the crossing is opened. In particular, the relevant information is not only the spectrum of $m$, but also the algebraic relation between the matrices $\gamma$ and $m$. The Dirac theory does not determine how this crossing connects to the rest of the Brillouin zone; that global information must come from the full lattice model or from continuity arguments. Nevertheless, it provides the simplest controlled setting in which to identify the possible complex dispersions that can in principle appear inside a gap.

\begin{figure*}[t]
    \centering
    \subfloat[Real momentum branches\label{fig: real momentum bands}]{
        \includegraphics[width=0.32\textwidth]{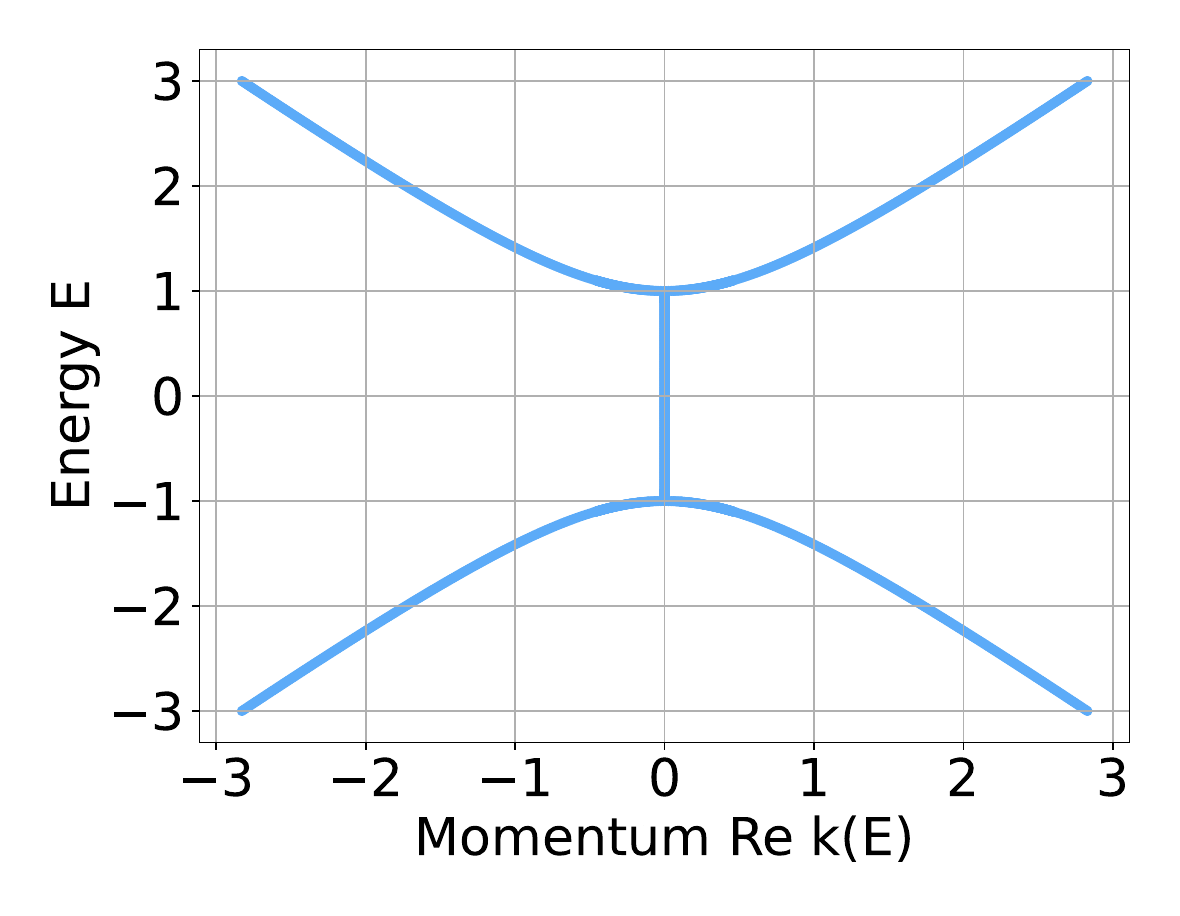}
    }\hfill
    \subfloat[Imaginary momentum branches\label{fig: imaginary momentum bands}]{
        \includegraphics[width=0.32\textwidth]{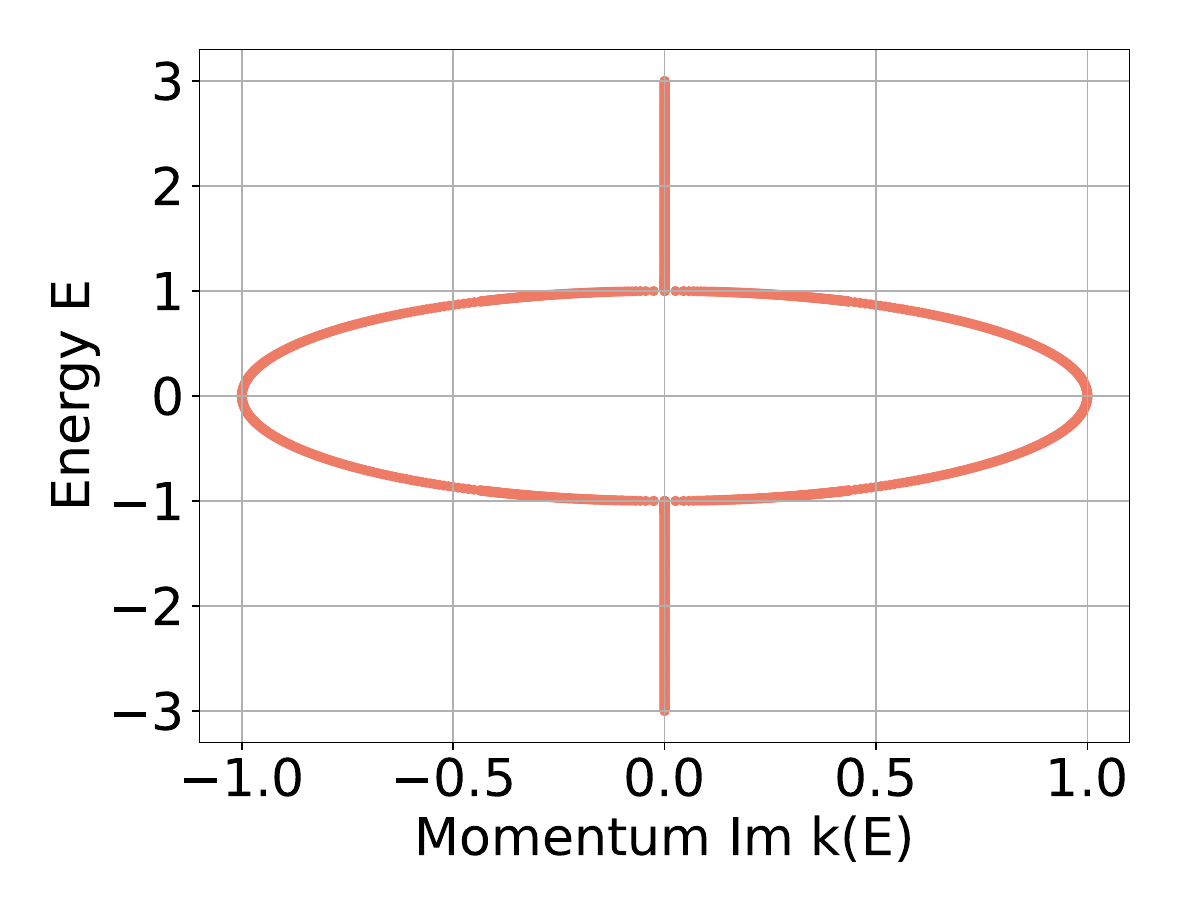}
    }\hfill
    \subfloat[Real and imaginary momentum branches\label{fig: combined real and imag}]{
        \includegraphics[width=0.32\textwidth]{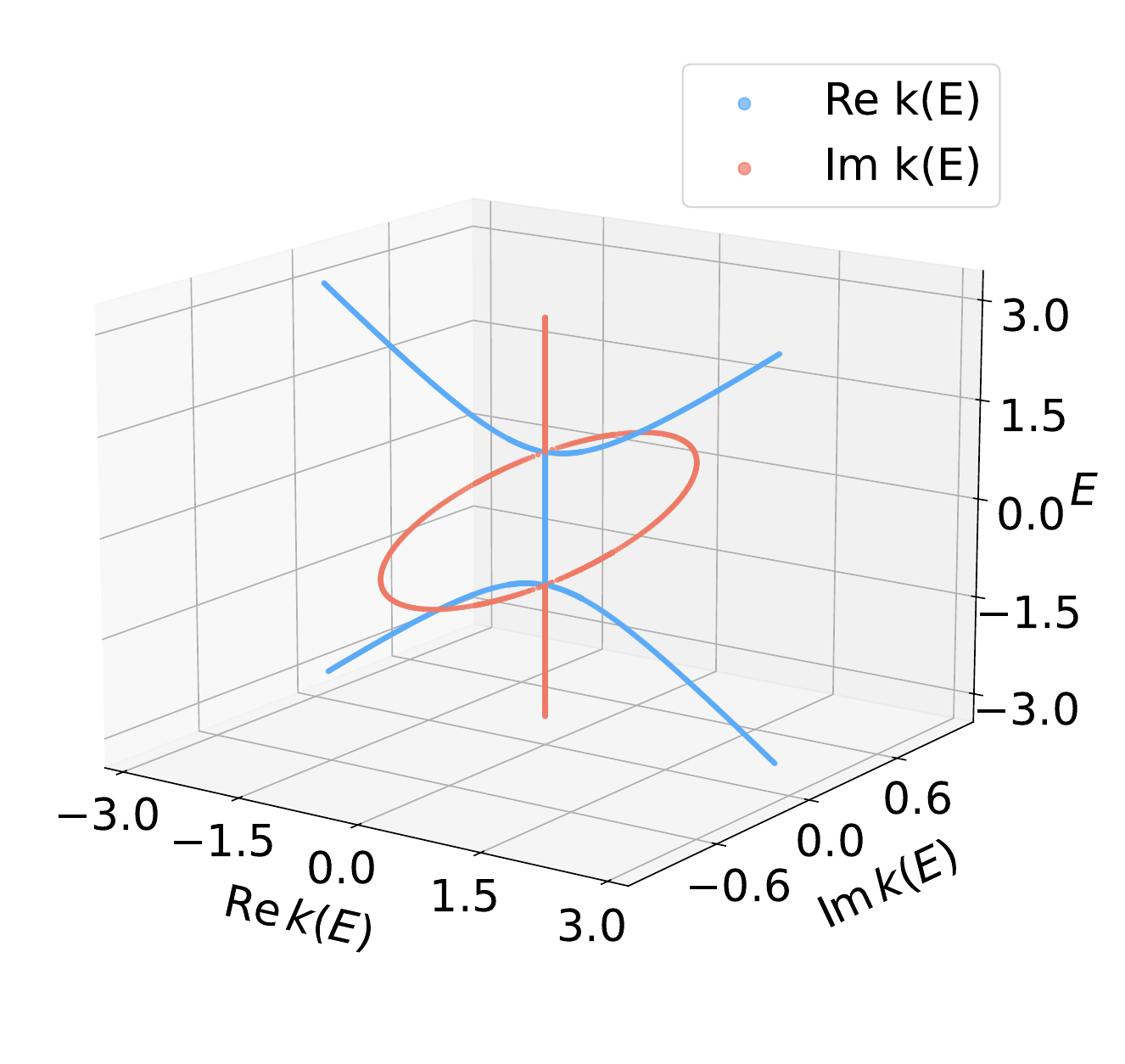}
    }
    \caption{Allowed momenta given by the Dirac momentum operator $\hat{k}(E) = \gamma^{-1}(E-m)$ as a function of $E$, with $\gamma=\sigma_y$ and $m=\sigma_x$. The blue (salmon) line represents the real (imaginary) part of the momentum.}
    \label{fig: 2 band model}
\end{figure*}

The transfer matrix of this long-wavelength theory follows from the real-space Dirac equation. To pass to real space, we perform the canonical replacement $p\rightarrow -\mathrm{i}\mathrm{d}/\mathrm{d}x$. The Schrödinger equation $H_{\mathrm{D}}\psi=E\psi$ becomes
\begin{equation}
    \left(-\mathrm{i}\gamma\frac{\mathrm{d}}{\mathrm{d}x}+m\right)\psi(x)=E\psi(x).
\end{equation}
This equation can be written as
\begin{equation}
    \frac{\mathrm{d}\psi}{\mathrm{d}x}(x)
    =
    \mathrm{i}\gamma^{-1}(E-m)\psi(x).
\end{equation}
Thus, propagation in space is generated by
\begin{equation}
    \psi(x+\Delta)=e^{\mathrm{i}\gamma^{-1}(E-m)\Delta}\psi(x),
    \qquad
    \hat{T}(E):=e^{\mathrm{i}\gamma^{-1}(E-m)}.
\end{equation}
In this sense, $\hat{T}(E)$ is the transfer matrix of the Dirac theory: it propagates the wavefunction in continuous real space by a unit of distance. Its eigenvalues are of the form $e^{\mathrm{i}k(E)}$, so that the operator $\hat{k}(E) = \gamma^{-1}(E-m)$ gives the allowed momenta as functions of energy. Real values of $k(E)$ describe propagating modes -- plane waves -- while complex values of $k(E)$ describe modes that grow or decay in space -- evanescent waves.

As a first example, we consider the minimal two-band Dirac Hamiltonian with $\gamma=\sigma_y$ and $m=\sigma_x$, where $\sigma_x$ and $\sigma_y$ are Pauli matrices. 
We show the momentum spectrum resulting from diagonalising $\hat{k}(E)$ in Fig.~\ref{fig: 2 band model}, where we plot $k_j(E)$ horizontally and $E$ vertically to conform with the usual convention for band structures. We call each $k_j(E)$ a momentum branch. Where an eigenvalue $k_j$ satisfies $\mathrm{Im}\,k_j(E)=0$, we call it a real-momentum branch; if $\mathrm{Re}\,k_j(E)=0$, we call it an imaginary-momentum branch. We combine the real and imaginary parts of the same momentum branches in the three-dimensional plot shown in Fig.~\ref{fig: combined real and imag}.

The 3D plot in Fig.~\ref{fig: combined real and imag} allows us to follow the momentum branches continuously as the energy is varied. With $\gamma=\sigma_y$ and $m=\sigma_x$, these branches are
\begin{equation}
    k_\pm(E)=\pm\sqrt{E^2-1}.
\end{equation}
For $|E|>1$, the two branches are real and reproduce the usual real-momentum spectrum. At the lower-energy band edge, $E=-1$, the two real-momentum branches meet at $k=0$. For $-1<E<1$, they continue as
\begin{equation}
    k_j(E)=\pm \mathrm{i}\sqrt{1-E^2},\quad j\in\{1,2\},
\end{equation}
and therefore lie on the imaginary-$k$ axis. At the upper-energy band edge, $E=1$, the reverse process occurs: the two imaginary-momentum branches meet again at $k=0$ and continue as real-momentum branches as energy is increased.

The coalescence points at $E=\pm1$ are exceptional points of the transfer matrix: we have
\begin{equation}
    \hat T(\pm1)=\mathbb{1}_{2\times 2}+N_\pm,
    \qquad
    N_\pm=\mathrm{i}\sigma_y(\pm \mathbb{1}_{2\times 2}-\sigma_x),
\end{equation}
with
\begin{equation}
    N_\pm^2=0,
    \qquad
    N_\pm\neq 0.
\end{equation}
Thus $\hat T(\pm1)-\mathbb{1}_{2\times 2}$ has a one-dimensional kernel: the two coincident eigenvalues share only one linearly independent eigenvector. This is the defining property of an exceptional point.
The explicit calculation above should not be viewed as a peculiarity of the chosen Pauli matrices. Intuitively, the two colliding branches belong to the same band, and as they merge, both corresponding transfer-matrix eigenvectors approach the one associated with the single Bloch state at the band extremum, so that they coalesce and the transfer matrix becomes defective. A general proof, formulated directly in terms of the transfer matrix, is given in Appendix~\ref{appendix: Existence of EP}. 

\begin{figure*}[t]
    \centering
    \subfloat[$|b|<|c|=0.55$\label{fig:no_eps}]{
        \includegraphics[width=0.32\textwidth]{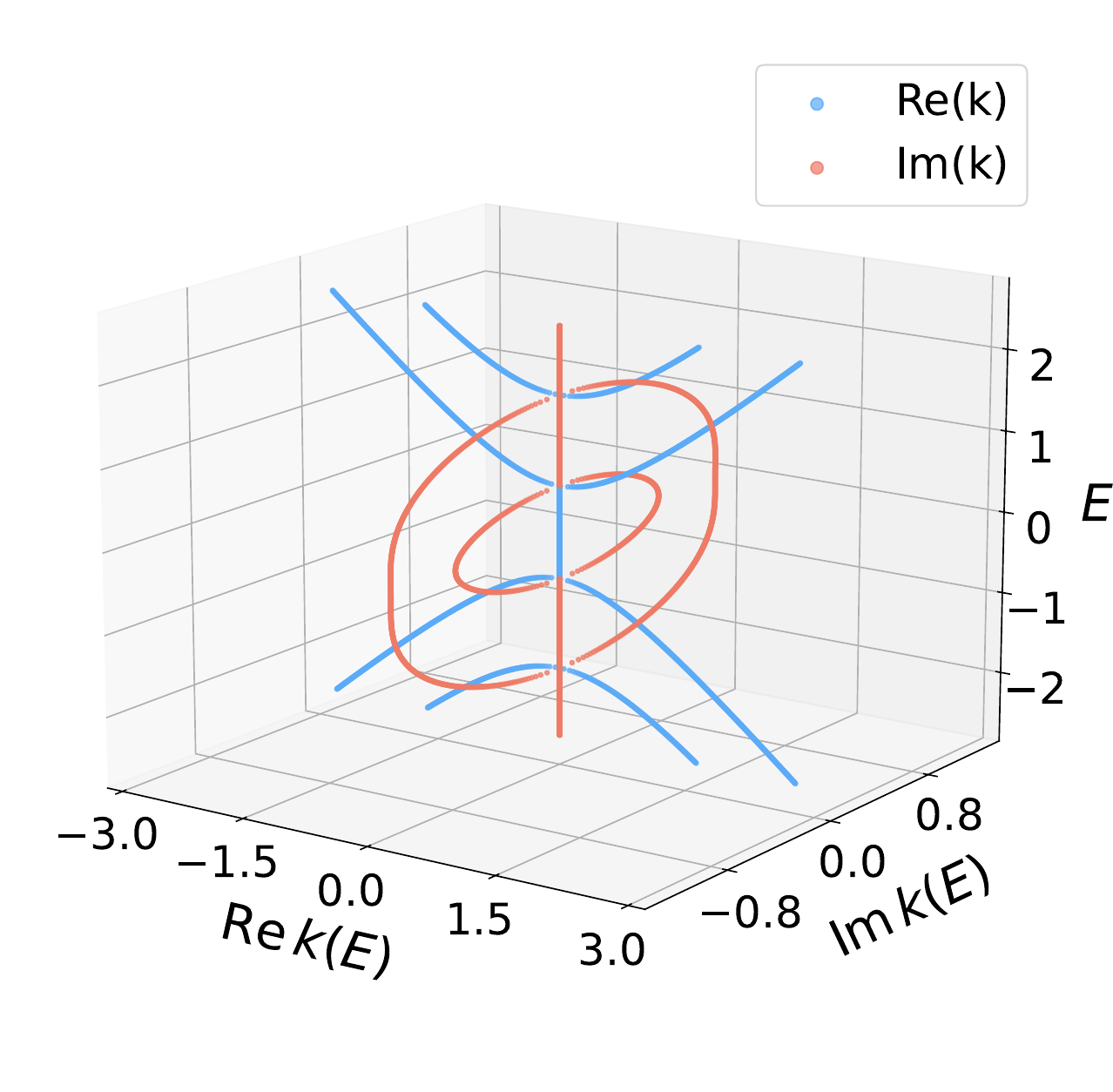}
    }\hfill
    \subfloat[$|b|=|c|=0.5$\label{fig: gap closing}]{
        \includegraphics[width=0.32\textwidth]{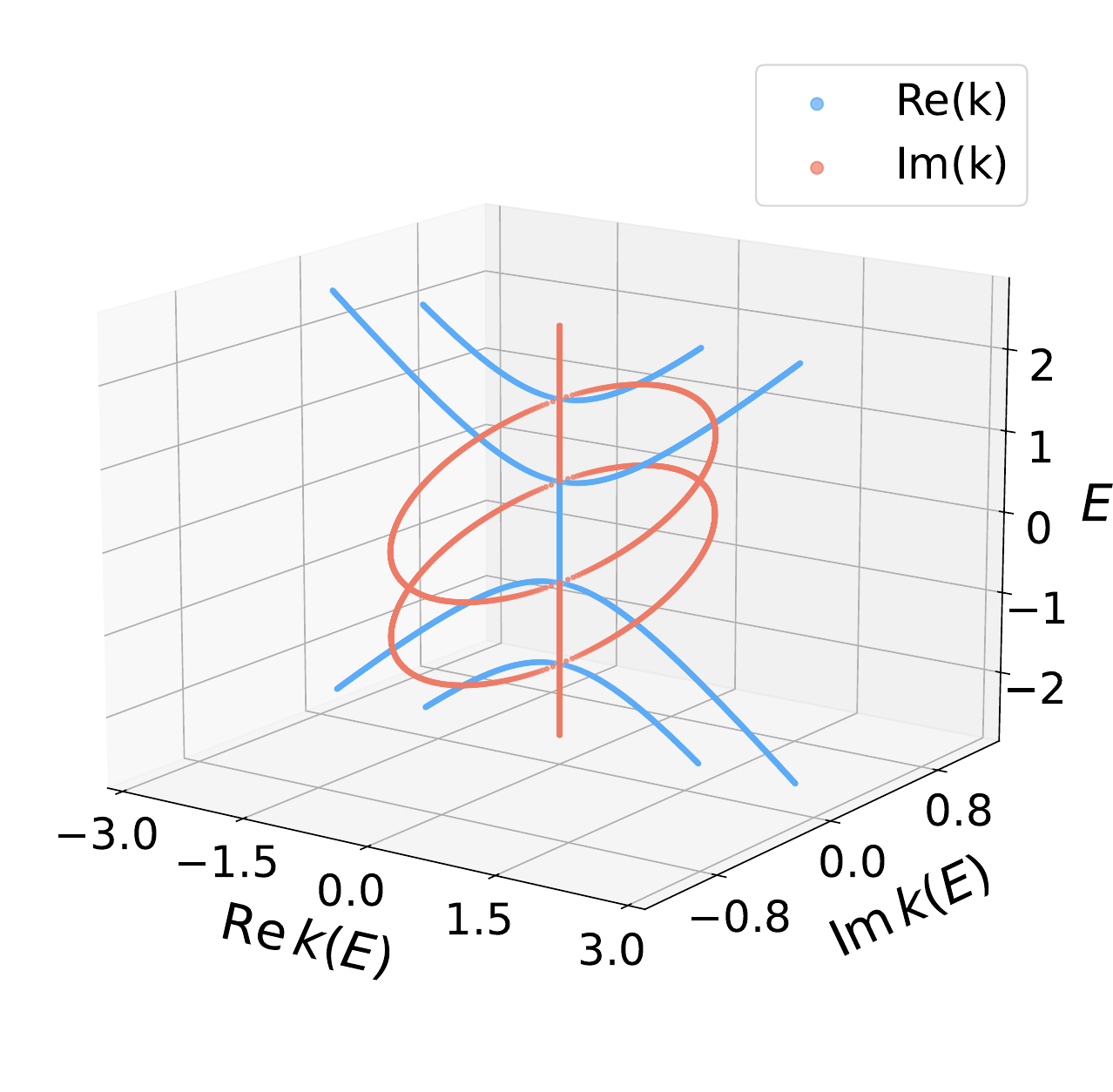}
    }\hfill
    \subfloat[$|b|>|c|=0.45$\label{fig:eps}]{
        \includegraphics[width=0.32\textwidth]{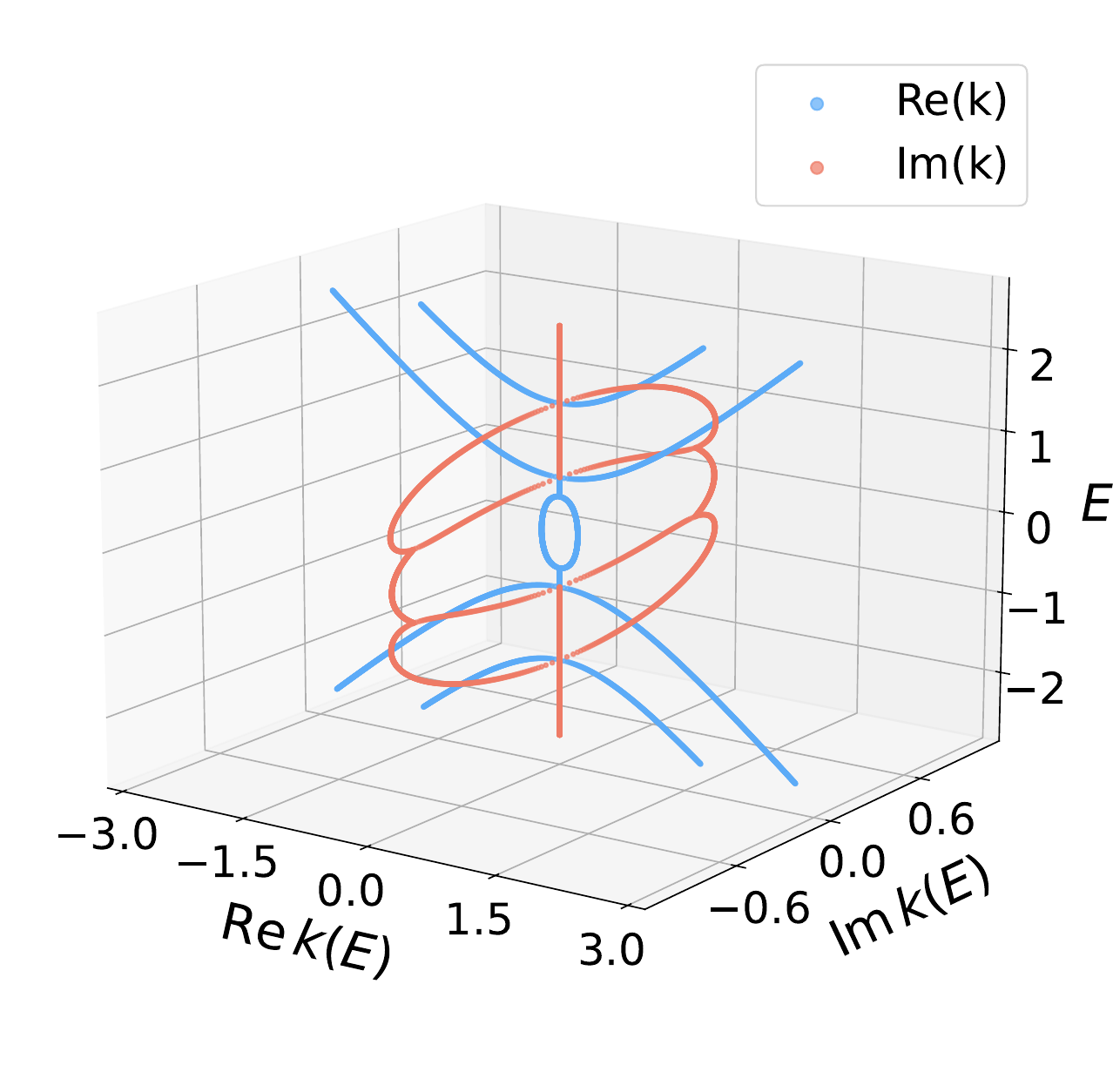}
    }

    \caption{
    Spectral phase transition between a phase without exceptional points inside the gap (a) and a phase with EPs inside the gap (c). The transition is mediated by a crossing between imaginary-momentum branches on the imaginary-$k$ axis, shown in (b). The model is given by Eq.~\eqref{eq: Dirac model with EPs} with the fixed values $a=1$ and $b=0.5$ and different values for $c$.}
    \label{fig:three_subfigs}
\end{figure*}

An interesting feature emerges when two imaginary-momentum branches coalesce: exceptional points appear inside the gap. To illustrate this, we next consider a four-band model constructed from two coupled two-band systems given by 
\begin{equation}
\label{eq: Dirac model with EPs}
    \begin{aligned}
        \gamma &= \mathbb{1}_2 \otimes \sigma_y,\\
        m &= a\mathbb{1}_2 \otimes \sigma_x + b\sigma_y \otimes \sigma_y + c\sigma_z \otimes \sigma_x,
    \end{aligned}
\end{equation}
 for different real values $a,b$ and $c$. In this construction, imaginary-momentum branches associated with different band edges can coalesce inside the gap. The corresponding spectrum is shown in Fig.~\ref{fig:three_subfigs} and reads 
\begin{equation}
\label{eq: analytic expression k(E) 4 bands}
\begin{aligned}
 k(E)=&\pm \sqrt{E^2-a^2+b^2-c^2\pm 2\Delta(E)},\\
 \Delta(E) \equiv&\sqrt{b^2(E^2-a^2)+a^2c^2}.
\end{aligned}
\end{equation}
Compared to the two-band model, the momentum spectrum now contains two pairs of imaginary-momentum branches, originating from two band edges. The separation between these two pairs is controlled by $\Delta(E)$. When $\Delta(E)\neq 0$, the two pairs are distinct. When $\Delta(E)=0$, the two choices of the inner sign in Eq.~\eqref{eq: analytic expression k(E) 4 bands} coincide, so different imaginary-momentum branches coalesce at the same energy. 

For $|b|<|c|$, $\Delta(E)$ does not vanish for real $E$, and the four imaginary-momentum branches remain separated throughout the gap, as shown in Fig.~\ref{fig:no_eps}. At the critical point $|b|=|c|$, the separation between the two pairs of imaginary-momentum branches closes at $E = 0$, as shown in Fig.~\ref{fig: gap closing}. For $|b|>|c|, a\neq 0 $, $\Delta(E)$ vanishes at two real energies,
\begin{equation}
    E_{\mathrm{EP}}^\pm
    =
    \pm |a|\sqrt{1-\frac{c^2}{b^2}}.
\end{equation}
These energies lie strictly inside the gap if and only if $b^4<c^2(a^2+b^2)$; otherwise the coalescence coincides exactly with the ordinary band-edge EPs. Indeed, minimising the lowest positive band $E_-^2(k)=a^2+b^2+c^2+k^2-2\sqrt{c^2(a^2+b^2)+b^2k^2}$ over $k$ gives an interior minimum at $k^2=\left[b^4-c^2(a^2+b^2)\right]/b^2$, which is admissible precisely when $b^4\geq c^2(a^2+b^2)$ and at which $E_-^2=a^2(1-c^2/b^2)=\left(E_{\mathrm{EP}}^{\pm}\right)^2$.

\begin{figure}[b]
    \centering
    \includegraphics[width=\linewidth]{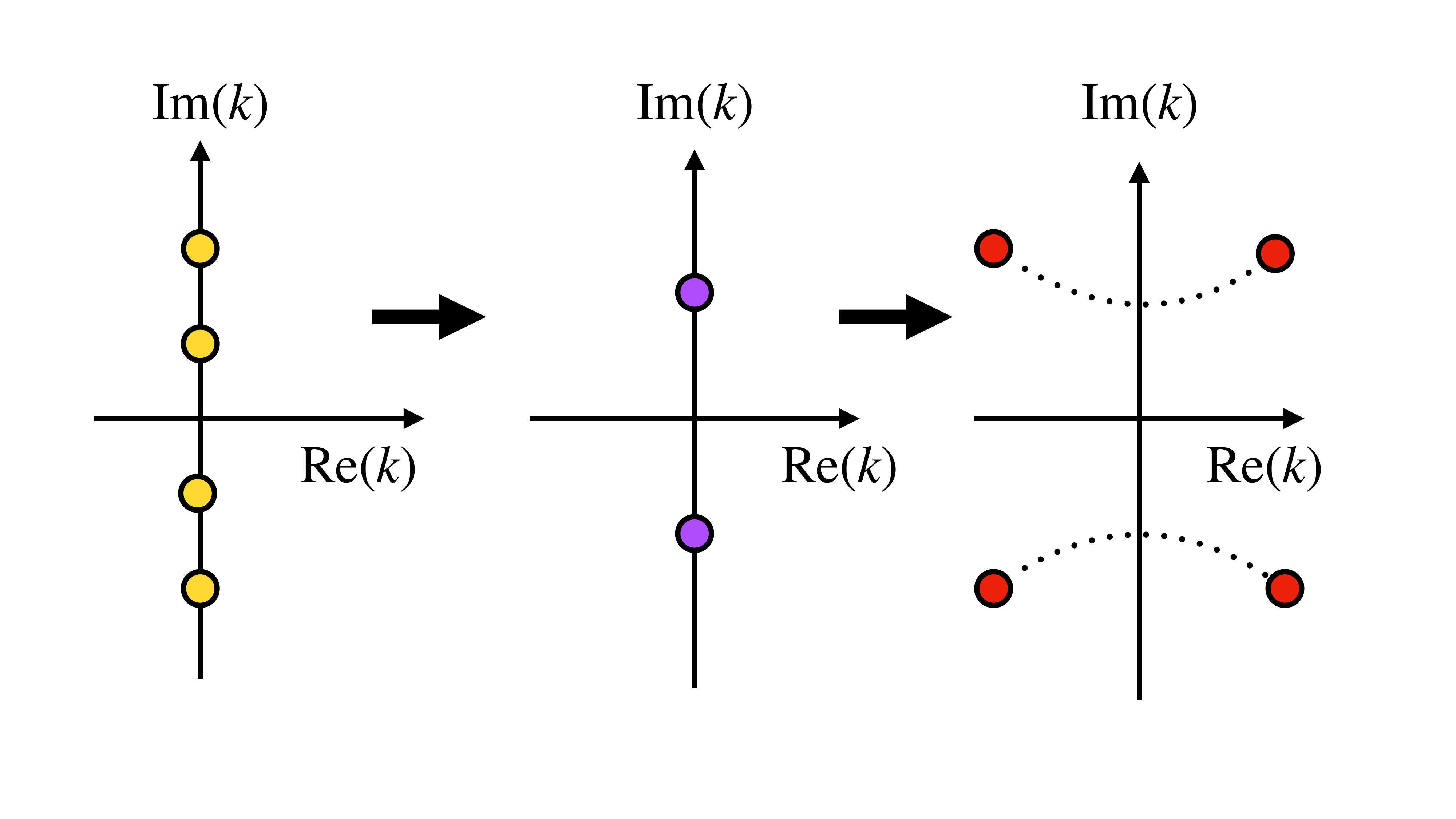}
    \caption{Illustration of the formation of in-gap EPs. The left panel shows the purely imaginary momentum solutions $k(E)$ (yellow circles), corresponding to the imaginary-momentum branches for energies between the band edge and the nearest EP. The middle panel depicts the coalescence of these imaginary momentum branches at an in-gap EP: the two decaying branches coalesce at one exceptional point of $T(E)$ and the two growing branches at another (purple circles). The right panel shows the splitting of the momentum solutions into the complex plane, yielding complex-valued $k(E)$ (red circles) for energies between the pair of in-gap EPs.}
    \label{fig: Illustration of the creation of EP}
\end{figure}

At energies given by $E_{\mathrm{EP}}^\pm$, imaginary-momentum branches coalesce. Between $E_{\mathrm{EP}}^-$ and $E_{\mathrm{EP}}^+$, the momenta are no longer purely imaginary but acquire both real and imaginary parts, as shown in Fig.~\ref{fig:eps} and illustrated schematically in Fig.~\ref{fig: Illustration of the creation of EP}.
The coalescence points at $E_{\mathrm{EP}}^\pm$ are the in-gap exceptional points. Their robustness relies on spinless TRS and a complex symplectic condition on the transfer matrix due to the Hermiticity of the underlying Hamiltonian. Appendix~\ref{appendix: Existence of EP} gives the corresponding eigenvector argument. In Sec.~\ref{sec: symmetry} we derive the symmetry constraints systematically and show in what sense these in-gap exceptional points are symmetry-protected.

We therefore identify two distinct gap structures in this model: an `insulating' phase of the evanescent wave spectrum without in-gap exceptional points for $|b|<|c|$, and a `semimetal' phase with a pair of in-gap exceptional points for $|b|>|c|$ with $a\neq 0$ and $b^4< c^2(a^2+b^2)$. The transition at $|b|=|c|$ is a crossing of the two pairs of imaginary-momentum branches, not a closing of the real-momentum energy bands.
The transition also has a topological interpretation. In Appendix~\ref{appendix: local topological index} we introduce a $\mathbb{Z}_2$ index $\nu(E)$ for the evanescent transfer-matrix spectrum, defined at every energy inside the gap at which no two transfer-matrix eigenvalues associated with decaying modes coincide. From now on, we refer to these eigenvalues as decaying eigenvalues. The index is the parity of the number of pairs related by TRS among the decaying transfer-matrix eigenvalues at that energy or, equivalently, the sign of a real discriminant. In all models studied here each in-gap energy carries at most two decaying modes, so that $\nu(E)=0$ when all of them decay purely exponentially and $\nu(E)=1$ when two of them form a pair related by TRS. Since $\nu$ can only change where two decaying eigenvalues coalesce, and does change at a generic (second-order) in-gap EP, the in-gap exceptional points are the domain walls of $\nu$. Non-generic coalescences at which $\nu$ does not change require fine tuning; the transition point $|b|=|c|$ itself is an example (Appendix~\ref{appendix: local topological index}). The gap in Fig.~\ref{fig:no_eps} is then trivial in the sense that $\nu(E)=0$ throughout, whereas in Fig.~\ref{fig:eps} the index equals one on the entire energy window bounded by the two in-gap EPs. In this sense the transition at $|b|=|c|$ is a topological transition of the evanescent spectrum, even though the real-momentum energy gap remains open. As for Weyl points in three dimensions, what is protected is each individual exceptional point, which cannot be removed on its own while it stays inside the gap; the $\nu=1$ window itself can be closed by annihilating the two EPs against each other. A second way for an in-gap EP to disappear from the gap is to reach a band edge, where $\nu$ ceases to be defined: in the present model this happens at $b^4=c^2(a^2+b^2)$, beyond which the EPs coincide with the band-edge EPs discussed above.

\section{Symmetry protection of in-gap exceptional points}
\label{sec: symmetry}

In the previous section, we used Dirac Hamiltonians as minimal models of gap openings and showed that their complex momentum spectra can contain exceptional points inside the gap. We now show that the spectral structure behind this phenomenon is not special to the Dirac approximation. Rather, it follows from general symmetry constraints on any 1D Hermitian Bloch Hamiltonian. 

We start by noting that the Bloch Hamiltonian is Hermitian for real $k\in \mathrm{BZ}$. If we extend $k$ to the complex plane, the Hermiticity condition becomes
\begin{equation}
\label{eq: (non)Hermiticity condition}
    H_{\mathrm{B}}^\dagger(k)=H_{\mathrm{B}}(k^*).
\end{equation}
In addition, we assume spinless TRS, corresponding to class AI in the Altland–Zirnbauer classification \cite{altland1997nonstandard}. For this class, we can take the time reversal operator as the complex conjugate operator without loss of generality. The TRS constraint can then be written as
\begin{equation}
    \label{eq: TRS condition}
    H_{\mathrm{B}}^*(k)=H_{\mathrm{B}}(-k^*).
\end{equation}
In particular, this condition is satisfied by both the $2 \times 2$ and the $4 \times 4 $ Dirac models that we have considered in Sec.~\ref{sec: in-gap transition}. 
The formal derivation of these conditions can be found in Appendix~\ref{appendix: props H_B in gap}.
We call $k$ an allowed momentum at real energy $E$ if
\begin{equation}
\label{eq: det equation allowed momenta}
    \det\!\left[H_{\mathrm B}(k)-E\right]=0.
\end{equation}
Equations~\eqref{eq: (non)Hermiticity condition} and~\eqref{eq: TRS condition} imply that this set of allowed momenta is closed under the maps
\begin{equation}
    k\mapsto k^*,
    \qquad
    k\mapsto -k^* .
\end{equation}
Thus, if $k$ solves the fixed-energy equation, then so do $k^*$ and $-k^*$. Applying the two operations successively also gives $-k$. Consequently, a generic complex momentum $k$ must appear in a quadruplet
\begin{equation} \label{eq: quadruplet}
    k, \qquad k^*,\qquad -k^*,\qquad -k.
\end{equation}
For $k=q+i\kappa$, with real $q,\kappa\neq0$, these are the four distinct momenta
\begin{equation}
    q+\mathrm{i}\kappa,\qquad q-\mathrm{i}\kappa,\qquad -q+\mathrm{i}\kappa,\qquad -q-\mathrm{i}\kappa .
\end{equation}
If $k$ is real, this quartet collapses to the pair $\pm k$. If $k$ is purely imaginary, it collapses to the pair $\pm \mathrm{i}\kappa$. Since $\mathrm{Re}\,k$ is defined $\bmod\,2\pi$, $k=\pi+\mathrm{i}\kappa$ also collapses the quartet to a pair of allowed momenta.

This symmetry constraint explains the qualitative spectral difference between the two- and four-band examples of the previous section. In the minimal two-band Dirac model, the transfer matrix has only two momentum branches. A generic four-element set of complex momenta cannot be accommodated, so the allowed momenta are restricted to one of the symmetry-invariant lines identified above: the real axis, the imaginary axis, or the line $\mathrm{Re}\,k=\pi$ (the last being absent in the continuum Dirac model, where $k$ is not periodic). The transition between these cases occurs when the two branches coalesce at a band edge.
In contrast, a transfer matrix with four momentum branches has enough room to realise the full symmetry-related quartet of complex momenta. Importantly, this is a condition on the number of momentum branches rather than on the number of energy bands: the condition is met by the four-band Dirac models considered here, but equally by a one-band lattice model with hopping range $R=2$, or a two-band lattice model with nearest-neighbour hoppings and an invertible (rank-2) hopping matrix [see the discussion around Eq.~\eqref{eq: NNH for supercells}].

We now want to understand these results from the point of view of the transfer matrix. Let $\Psi_n$ denote the vector of wavefunction components in the $n$th supercell, as in Eq.~\eqref{psin def}. In a translationally invariant bulk, we can make the following ansatz:
\begin{equation}
    \Psi_{n+1}=z\Psi_n.
\end{equation}
We refer to $z$ as the \textit{multiplier} of the solution: the factor by which it is rescaled from one supercell to the next. Writing $z=e^{\mathrm{i}k}$ then gives the same complex momentum that appears in Eq.~\eqref{eq: det equation allowed momenta}.
The eigenvalues of the transfer matrix are $z_j(E)=e^{\mathrm{i}k_j(E)}$, as shown in Appendix~\ref{appendix: spatial multipliers}. Thus the Bloch-Hamiltonian and transfer-matrix formulations describe the same allowed momenta, although their eigenvectors need not coincide.

Next, we show that the in-gap collisions are not just degeneracies but EPs of the transfer matrix. Since the TRS constraint implies that $T(E)$ can be chosen real (Appendix~\ref{appendix: Existence of EP}), $T(E)\phi=e^{\mathrm{i}k}\phi$ implies $T(E)\phi^*=e^{-\mathrm{i}k^*}\phi^*$, so that complex conjugation $\mathcal{K}$ maps
\begin{equation}
\mathcal{K}:\ V_k(E)\longrightarrow V_{-k^*}(E),\qquad
V_k(E):=\ker\!\left[T(E)-e^{\mathrm{i}k}\right].
\end{equation}
For purely imaginary momenta, $k=\mathrm{i}\kappa$, one has $-k^*=k$, and $\mathcal{K}$ maps each nondegenerate eigenspace onto itself. Once two imaginary branches have collided and moved into the complex plane as $k_\pm=\pm q+\mathrm{i}\kappa$ with $q\neq0$, one instead has $-k_+^*=k_-$, and $\mathcal{K}$ exchanges the two eigenspaces. Let $\phi_1(E)$ and $\phi_2(E)$ be the eigenvectors of the two colliding branches, which are nondegenerate away from the collision energy $E_{\mathrm{EP}}$ and can be followed continuously through it. Approaching $E_{\mathrm{EP}}$ from the imaginary side gives $\mathcal{K}\phi_1(E_{\mathrm{EP}})\propto\phi_1(E_{\mathrm{EP}})$, whereas approaching it from the complex side gives $\mathcal{K}\phi_1(E_{\mathrm{EP}})\propto\phi_2(E_{\mathrm{EP}})$. At $E_{\mathrm{EP}}$, $\mathcal{K}$ must therefore map the eigenvector onto itself and onto its partner simultaneously, which is possible only if
\begin{equation}
\phi_1(E_{\mathrm{EP}})\propto\phi_2(E_{\mathrm{EP}}),
\end{equation}
i.e., the two eigenvectors coalesce. In Appendix~\ref{app: defectiveness of T} we furthermore construct an explicit Jordan chain, showing rigorously that $T(E_{\mathrm{EP}})$ is indeed defective and that an isolated two-branch collision is a second-order EP.

Additionally, from the Hermiticity of the Hamiltonian, it is possible to show that the transfer matrix obeys the complex symplectic condition (also in Appendix~\ref{appendix: Existence of EP})
\begin{equation}
    \label{eq: symplectic condition}
    T(E)^{\dagger}\, \mathcal{F}\, T(E)=\mathcal{F},\quad \mathcal{F}=\begin{pmatrix}
        0&\Sigma\\
        -\Sigma &0
    \end{pmatrix},
\end{equation}
where $\Sigma$ is given by the reduced SVD, $J=U\Sigma W^{\dagger}$ with $J$ defined in Eq.~\eqref{eq: NNH for supercells}. This condition can be seen as a similarity transformation 
\begin{equation}
    T(E)^\dagger=\mathcal{F} \, T(E)^{-1} \, \mathcal{F}^{-1},
\end{equation}
so that for every allowed momentum $k$ we must also have $k^*$. Together with TRS we again obtain the quadruplet in Eq.~\eqref{eq: quadruplet}.

While the band edges harbour transfer-matrix EPs that follow from Hermiticity alone (Appendix~\ref{app: defectiveness of T}), our main focus is on the in-gap EPs discussed earlier, which are protected by the TRS constraint. Under small TRS-preserving perturbations, these may move along the energy axis, but they cannot be removed individually while they remain inside the gap: they can disappear only by pairwise annihilation, or by being pushed into a band edge. This robustness is topological in origin. Following the classification of Ref.~\cite{Kawabata2019PRX}, we regard the transfer matrix $T(E)$ as a non-Hermitian matrix parametrised by the energy $E$. Spinless TRS places $T(E)$ in class AI of the 38 non-Hermitian symmetry classes. The eigenvalues of a real matrix are either real or come in complex-conjugate pairs. Generically, the two eigenvalues that coalesce at an in-gap EP do so on the real axis, at $k_{\mathrm{EP}}\in\mathrm{i}\mathbb{R}$ or, for negative multipliers, at $k_{\mathrm{EP}}\in\pi+\mathrm{i}\mathbb{R}$, where two real transfer-matrix eigenvalues turn into a complex-conjugate pair. The reference point $e^{\mathrm{i}k_{\mathrm{EP}}}$ is then real, so that $\det[T(E)-e^{\mathrm{i}k_{\mathrm{EP}}}\mathbb{1}]$ is real, and whenever $T(E)$ has a \emph{point gap} (\emph{i.e.}, no eigenvalues) at the reference point $e^{\mathrm{i}k_{\mathrm{EP}}}$, i.e., $\det[T(E)-e^{\mathrm{i}k_{\mathrm{EP}}}\mathbb{1}]\neq 0$, its sign defines the zero-dimensional $\mathbb{Z}_2$ invariant of class AI~\cite{Kawabata2019PRX},
\begin{equation}
    (-1)^{\nu_{\mathrm{pg}}(E)}=-\,\mathrm{sgn}\det\!\left[e^{\mathrm{i}k_{\mathrm{EP}}}\mathbb{1} - T(E)\right].
    \label{eq: Z2 index}
\end{equation}
Here $k_{\mathrm{EP}}$ is the collision momentum of the EP under consideration; when different EPs collide at different momenta, as in the photonic crystal of Sec.~\ref{sec: photonic crystal}, Eq.~\eqref{eq: Z2 index} is evaluated near each EP with its own reference point, and the global labelling follows from the discriminant of Appendix~\ref{appendix: local topological index}.
The invariant $\nu_{\mathrm{pg}}(E) = 0,1 \in \mathbb{Z}_2$ can only change where the point gap closes. Near the EP under consideration, this happens only at the EP itself, where the determinant vanishes and changes sign; further away, the point gap can also close accidentally, when a single real multiplier crosses the reference point. A single EP therefore cannot be removed by a weak symmetry-preserving perturbation, since the surrounding point gap fixes $\nu_{\mathrm{pg}}$ on both sides of it. The global labelling is given by the index $\nu(E)$, the parity of the number of complex-conjugate pairs among the decaying multipliers: near each EP it jumps exactly where $\nu_{\mathrm{pg}}$ does, but unlike the latter it changes at every generic in-gap EP and nowhere else. A pair of in-gap EPs at $E_{\mathrm{EP}}^{(1)}<E_{\mathrm{EP}}^{(2)}$ thus divides the gap into three regions, with $\nu(E)=1$ for $E_{\mathrm{EP}}^{(1)}<E<E_{\mathrm{EP}}^{(2)}$ and $\nu(E)=0$ on either side: the EPs are domain walls of $\nu$ along the energy axis. Inside the gap, two EPs can disappear only by merging, which eliminates the $\nu=1$ region between them. The only other way for an in-gap EP to disappear from the gap is through a band edge, where the decaying multipliers reach the unit circle and $\nu$ is no longer defined. This happens in the Dirac model of Eq.~\eqref{eq: Dirac model with EPs} at $b^4=c^2(a^2+b^2)$.

\begin{figure}[t]
    \centering
    \includegraphics[width=0.8\linewidth]{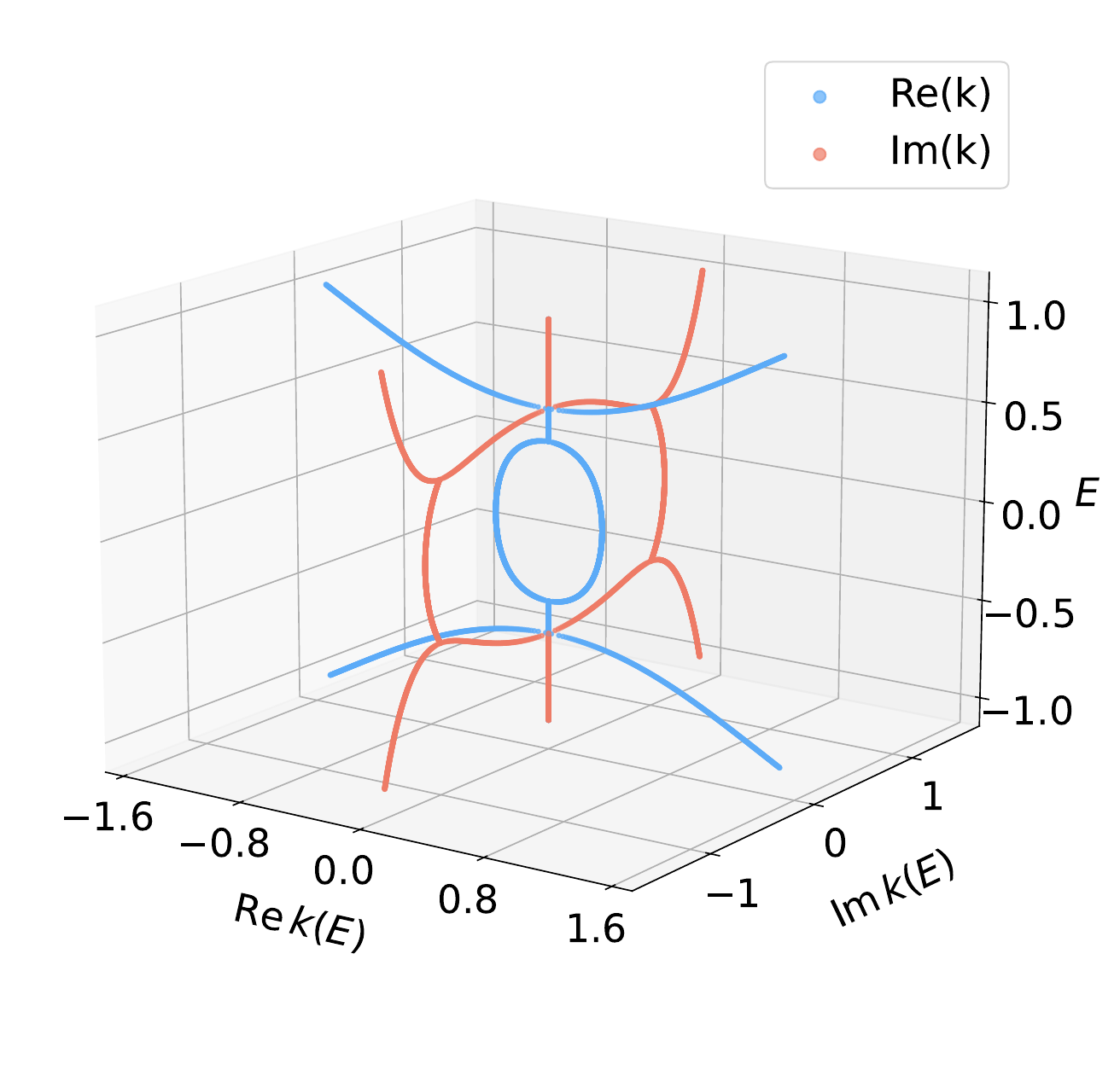}
    \caption{Complex momenta $k=-\mathrm{i}\log\lambda$ obtained from the bulk transfer-matrix spectrum $\{\lambda\}$ of the tight-binding model of Eq.~\eqref{eq: tight-binding model with EPs}, for $a=1.275$, $b=-3.7$, $c=3$ and $j_0=1$, zoomed into the central gap. The band edges lie at $E\simeq\pm0.560$ with $k_0=0$, and the in-gap exceptional points at $E\simeq\pm0.398$.}
    \label{fig: tight-binding model with EPs}
\end{figure}

\section{Physical consequences}
\label{sec: physical consequences}
The in-gap exceptional points produce a sharp real-space signature: they change the asymptotic tails of localised bound states from purely exponential to oscillatory. The distinction is controlled by the leading evanescent momentum at the bound-state energy. If this momentum is purely imaginary, the tail decays without oscillations; if it has a nonzero real part, as occurs between a pair of in-gap EPs, the tail becomes an exponentially damped oscillation. Thus the transfer-matrix spectrum reveals physical information about the gap that is not captured by a conventional expansion around the nearest energy band edge.

\begin{figure*}[t]
    \centering
    \subfloat[\label{fig: No Oscillations}]{
        \includegraphics[width=0.4\textwidth]{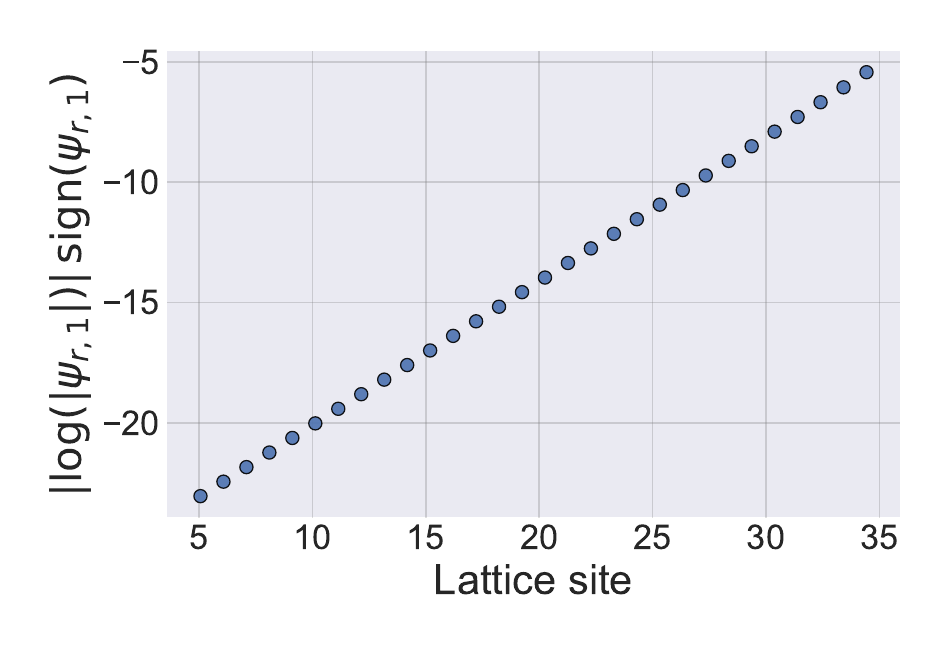}
    }
    \subfloat[\label{fig: Oscillations}]{
        \includegraphics[width=0.4\textwidth]{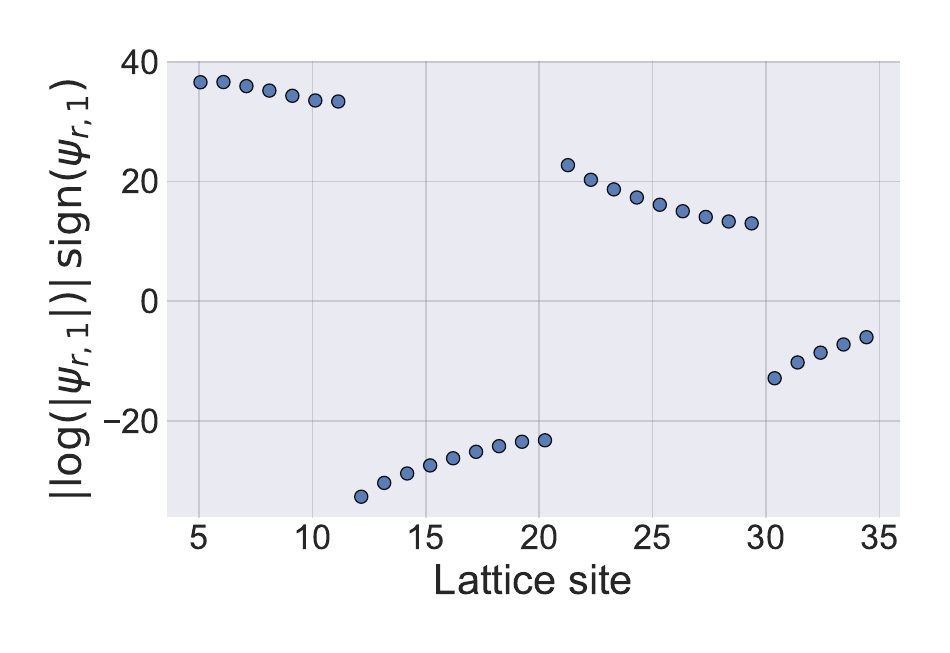}
    }

    \caption{
    Left portion of $\chi(\psi_{r,1})$, showing bulk behaviour away from both the boundaries and the central local potential. The bound state $\psi_{r}$ is constructed such that: (a) its energy lies between the band edge and the closest exceptional point, $E_{\mathrm{b}}\simeq-0.479$ for $U_0=0.5$, and (b) its energy lies between the pair of in-gap exceptional points, $E_{\mathrm{b}}\simeq-0.125$ for $U_0=8$. The lattice has $N=80$ unit cells.}
    \label{fig: Bound state}
\end{figure*}

To illustrate this, we consider a 1D tight-binding Hamiltonian whose transfer matrix exhibits EPs within the gap. In principle, one could consider a model whose low-$k$ limit is described by a Dirac Hamiltonian $H_{\mathrm{D}}$ that hosts in-gap EPs, as previously shown. However, this description is restricted to the vicinity of $k=0$, where the gap is small and the real part of the complex momentum remains correspondingly small. As a result, the associated oscillations occur on large length scales and are not clearly resolved in finite size calculations. To obtain a clearer picture, we instead consider the following tight-binding model, which exhibits EPs within the gap without relying on a low-energy approximation. It is a four-orbital chain with open boundary conditions, and its parameters $a$, $b$, $c$ are unrelated to those of the Dirac model of Eq.~\eqref{eq: Dirac model with EPs}:
\begin{equation}
    \begin{aligned}
        \label{eq: tight-binding model with EPs}
        H=\sum_{r=1}^{N}\sum_{\alpha,\beta=1}^{4} \bigg[&M_{\alpha,\beta}\ket{r,\alpha}\bra{r,\beta}\\
        &+\big(J_{\alpha,\beta}\ket{r,\alpha}\bra{r+1,\beta}+\mathrm{h.c.}\big) \bigg],
    \end{aligned}
\end{equation}
\begin{equation}
    \begin{aligned}
        M&=a\sigma_z\otimes\sigma_z+b\sigma_y\otimes\sigma_y+c\sigma_x\otimes\sigma_z,\\
        J&=j_0\begin{pmatrix}
        0&0&0&0\\
        0&0&0&0\\
        0&1&0&0\\
        -1&0&0&0
    \end{pmatrix}. 
    \end{aligned}
\end{equation}
with $a,b,c,j_0\in\mathbb{R}$ and $N$ an even number, where the hopping term is omitted for $r=N$. Both $M$ and $J$ are real, so the model lies in class AI, and $J^2=0$ with $\mathrm{rank}\,J=2$, so the transfer matrix is $4\times4$ and has four momentum branches. Its spectrum is shown in Fig.~\ref{fig: tight-binding model with EPs}, where we zoom in to focus on the gap. We then introduce a local scalar potential at the centre of the lattice by 
\begin{equation} \label{eq: with potential}
    H\rightarrow H+U_0\sum_{\alpha=1}^{4}\ket{N/2,\alpha}\bra{N/2,\alpha},
\end{equation}
for different values of the scalar $U_0\in\mathbb{R}$ (the derivation in Appendix~\ref{appendix: Asymptotic tails} allows for a general matrix-valued onsite potential), which stabilises a bound state with energy inside the gap.
From a Taylor expansion of the band dispersion around the band edge, one would expect that the bulk behaviour of this bound state does not exhibit oscillations, since the band edges occur at $k_0 = 0$, implying purely exponential decay. However, in the following we show that the validity of this conclusion depends on the energy of the bound state.
To be precise, we now connect the in-gap transfer-matrix spectrum to the spatial profile of a localised bound state. Consider a local potential at $r_0$ producing a bound state of energy $E_{\mathrm{b}}$ within a bulk gap. Outside the finite region where the potential acts, the wavefunction obeys the homogeneous bulk Schrödinger equation. Consequently, the asymptotic tail is a linear combination of the transfer-matrix eigenmodes evaluated at $E_{\mathrm{b}}$. Since the bound state must decay away from the impurity, only modes with $|e^{\mathrm{i}k({E_\mathrm{b}})}|<1$ contribute to the right tail, whereas only modes with $|e^{\mathrm{i}k(E_{\mathrm{b}})}|>1$ contribute to the left tail. The coefficients of these modes are fixed by matching conditions through the impurity region, while the allowed complex momenta are determined entirely by the bulk transfer-matrix spectrum. This decomposition holds exactly in the thermodynamic limit, as required by normalisation; in a finite system, we therefore focus on the bulk region, away from the boundaries (Appendix~\ref{appendix: Asymptotic tails}).

Asymptotically, the bulk components of a generic bound state are governed by the slowest-decaying mode among the allowed transfer-matrix eigenmodes. Writing the slowest-decaying momentum as $k(E_{\mathrm{b}})=q_r+\mathrm{i}q_i$, with $q_i>0$ for the right tail, the corresponding contribution behaves as $e^{-q_i(r-r_0)}$. Hence, the asymptotic right tail is oscillatory precisely when the leading decaying mode has ${\rm Re}(k)\neq 0,\pm\pi$. If the slowest-decaying mode has purely imaginary momentum, the tail is purely exponential. An identical argument applies to the left tail. At an in-gap EP, the bound state switches between these two types of decay. Exactly at the EP, the two leading decaying modes coalesce and $T(E_{\mathrm{b}})$ is defective, so the tail is neither a single exponential nor a damped standing wave: far from the impurity, every component of the right tail behaves generically as $(r-r_0)e^{\mathrm{i}k_{\mathrm{EP}}(r-r_0)}$, and the left tail analogously. A detailed derivation is presented in Appendix \ref{appendix: Asymptotic tails}.
 
For numerical verification, we obtain the spatial profile of the mid-gap bound state of Eq.~\eqref{eq: with potential} that appears for a wide range of $U_0 > 0$, and only consider its wavefunction component for the orbital $\alpha=1$ for simplicity, giving $\psi_{r,1}$ (which can be chosen to be real). We then compute 
\begin{equation}
    \chi(\psi_{r,1}):=|\log(|\psi_{r,1}|)|\, \mathrm{sign}(\psi_{r,1}).
\end{equation}
Since $|\psi_{r,1}| < 1$, the magnitude $|\log|\psi_{r,1}||$ grows linearly with the distance from the impurity, with a slope set by the decay rate $q_i$, while the factor $\mathrm{sign}(\psi_{r,1})$ reflects the curve to negative values wherever the wavefunction changes sign. A purely exponential tail therefore appears as a straight line of constant sign, whereas an oscillatory tail appears as a line whose sign alternates, with jumps at the nodes of $\psi_{r,1}$, roughly every $\pi/q_r$ sites. The energy of the bound state is controlled by tuning the parameter $U_0$. If the energy lies between the band edge and the nearest EP, the decay is indeed purely exponential, as shown in Fig.~\ref{fig: No Oscillations}. In contrast, if the bound state energy lies between the two EPs, the allowed momentum becomes complex with a nonzero real part, leading to oscillatory behaviour in the bulk, as shown in Fig.~\ref{fig: Oscillations}. 

\section{Photonic crystal realisation of in-gap exceptional points}
\label{sec: photonic crystal}

In the previous section, we examined the physical consequences of in-gap EPs in a lattice tight-binding model. Although this analysis is specific to a lattice system, the key ingredients underlying EPs within the gap are not. Here, we demonstrate this by providing an explicit realisation based on an anisotropic photonic crystal (PhC), whose transfer matrix exhibits the same essential features: the reality condition and the symplectic condition.

A photonic crystal is a dielectric material whose relative permittivity $\epsilon$ changes periodically in space. Here we consider a material that is periodic in the $z$-direction only, $\epsilon(z + a) = \epsilon(z)$, where $a$ is the period and $z$ is the spatial coordinate in which $\epsilon$ changes. For simplicity, we model $\epsilon$ as a piecewise-constant function taking two distinct matrix-values over a single period and we work in dimensionless units using $c=\mu_0=\epsilon_0=1$:
\begin{equation}
\epsilon(z) = \begin{pmatrix}
        \epsilon_{xx} & \epsilon_{xy} & \epsilon_{xz}\\
        \epsilon_{yx} & \epsilon_{yy} & \epsilon_{yz}\\
        \epsilon_{zx} & \epsilon_{zy} & \epsilon_{zz}
        \end{pmatrix}=\begin{cases}
\epsilon_1 \quad & 0 \leq z < a_1, \\
\epsilon_2 \quad & a_1 \leq z < a_1+a_2 = a.
\end{cases}
\end{equation}
We consider illumination by a plane wave propagating along $z$ with frequency $\omega$. Inside the crystal the displacement field solution to the Maxwell equations then takes the form (Appendix~\ref{appendix: construction of T for PC})
\begin{equation}
\begin{aligned}
    D(z,t) &=
    \begin{pmatrix}
        \tilde{\epsilon}E_{2\mathrm{D}}(z,t)\\
        0
    \end{pmatrix},
\end{aligned}
\end{equation}
where 
\begin{equation}
\begin{aligned}
    E_{2\mathrm{D}}(z,t) &= e^{-\mathrm{i}\omega t}\Bigg[
    v_1
    \left(c_1e^{\mathrm{i}k_1 z}+c_2e^{-\mathrm{i}k_1 z}\right) \\
    &\qquad\qquad
    +
    v_2
    \left(d_1e^{\mathrm{i}k_2 z}+d_2e^{-\mathrm{i}k_2 z}\right)
    \Bigg]
\end{aligned}
\end{equation}
gives the $x-y$ components and $v_1$ and $v_2$ are the eigenvectors of the matrix $\tilde{\epsilon}$ given by
\begin{equation}
    \tilde{\epsilon}=\begin{pmatrix}
        \epsilon_{xx}-\epsilon_{xz}\epsilon_{zx}/\epsilon_{zz} & \epsilon_{xy}-\epsilon_{xz}\epsilon_{zy}/\epsilon_{zz}\\
        \epsilon_{yx}-\epsilon_{yz}\epsilon_{zx}/\epsilon_{zz} & \epsilon_{yy}-\epsilon_{yz}\epsilon_{zy}/\epsilon_{zz}
    \end{pmatrix},
\end{equation}
with the eigenvalues $k_1^2/\omega^2$ and $k_2^2/\omega^2$, respectively. From the continuity of the in-plane electric and magnetic fields and of the normal component of the displacement field, imposed by Maxwell's equations, we can obtain (in Appendix~\ref{appendix: construction of T for PC}) an eigenvalue equation for the corresponding transfer matrix:
\begin{equation}
        \mathcal{T}(\omega) \Psi_1(\omega)=e^{\mathrm{i}k(\omega)a}\Psi_1(\omega),\quad
        \Psi_1(\omega):=\begin{pmatrix}
            c_{1,1}\\c_{2,1}\\d_{1,1}\\d_{2,1}
        \end{pmatrix},
\end{equation}
where the subscript $1$ means the coefficients with respect to the first layer of the PhC. This transfer matrix propagates the coefficients of the in-plane electric field by a full period. Its eigenvalues are $e^{\mathrm{i}k(\omega)a}$ and therefore determine the crystal momentum as a function of frequency, $k(\omega)$.

\begin{figure}
    \centering
    \includegraphics[width=0.8\linewidth]{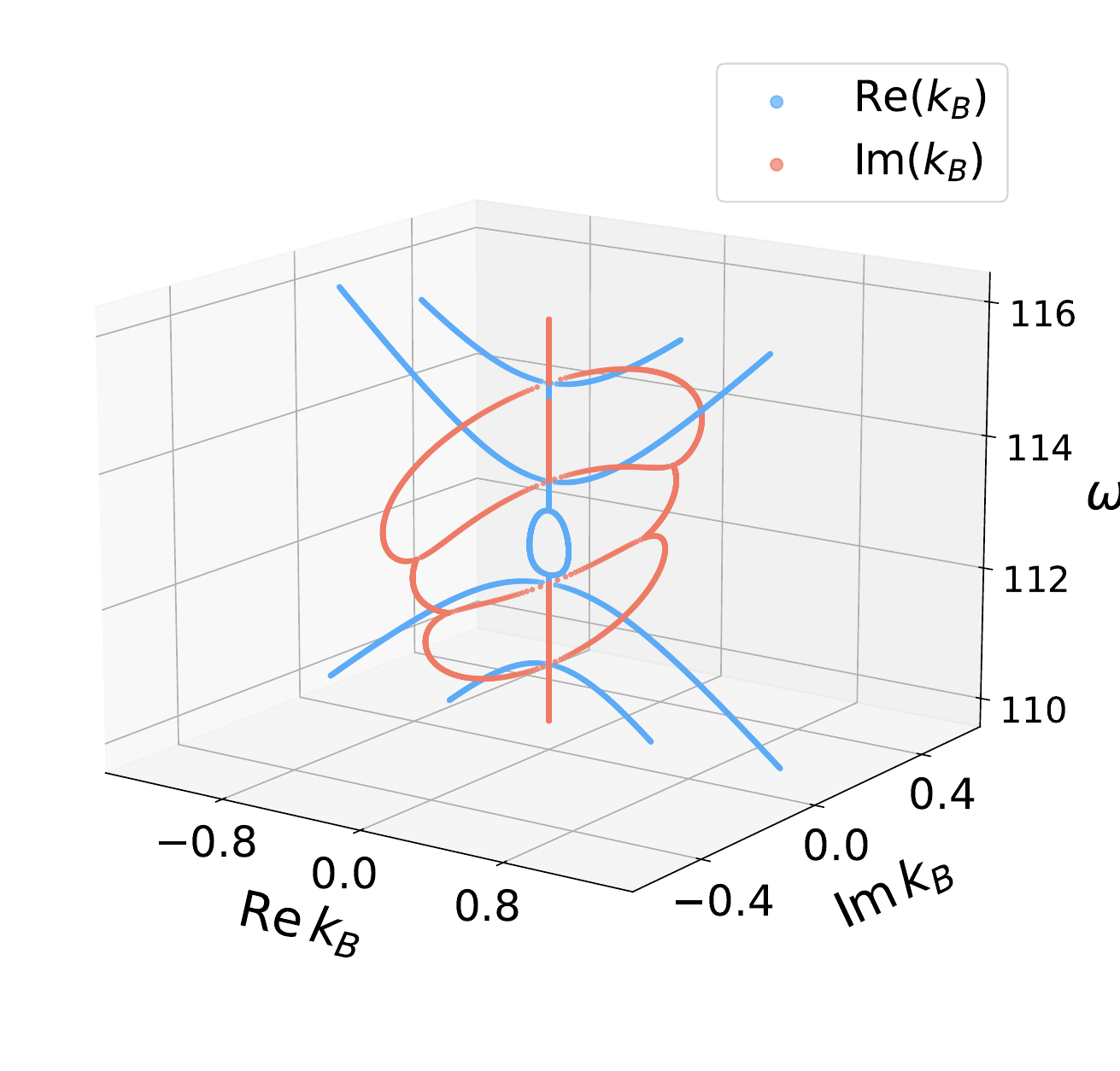}
    \caption{Allowed momenta $k(\omega)$ obtained from the spectrum of the transfer matrix $\mathcal{T}(\omega)$ of the anisotropic photonic crystal defined by Eq.~\eqref{eq: epsilon PC}, in a portion of a gap. As in the lattice models, two evanescent branches leave the imaginary-momentum axis at a pair of in-gap exceptional points and acquire opposite real parts in between.}
    \label{fig: EPS in PC}
\end{figure}

Since Maxwell's equations have real coefficients, if $E_{2\mathrm{D}}$ is a solution then $E_{2\mathrm{D}}^*$ is also a solution. From this fact, we derive (shown in Appendix~\ref{appendix: properties of Y for PC}) that the transfer matrix of the PhC possesses an antiunitary symmetry, i.e. there is an antilinear operator $P$ such that
\begin{equation}
    \begin{aligned}
        P\mathcal{T}(\omega) P=\mathcal{T}(\omega),\quad P^2=\mathbb{1},\quad \forall\omega\in\mathbb{R}.
    \end{aligned}
\end{equation}
Moreover, the transfer matrix satisfies a symplectic condition, which follows from an underlying Hamiltonian structure (see Appendix~\ref{appendix: properties of Y for PC}). Consequently,
\begin{equation}
\mathcal{T}(\omega)\sim \mathcal{T}(\omega)^{-1},\quad\forall\omega\in\mathbb{R}.
\end{equation}
As a result, $\mathcal{T}(\omega)$ exhibits the same structural properties as the transfer matrices considered previously, and thus admits the same spectral features. To support this conclusion, we present a portion of the band structure of a PhC given by
\begin{equation}
\label{eq: epsilon PC}
    \tilde{\epsilon}(z)=\begin{cases}
        w_1\mathbb{1}_2+w_2\sigma_x\quad\quad\quad\quad\quad\; 0\leq z< a_1,\\
        u_1\mathbb{1}_2+u_2\sigma_x+u_3\sigma_z\quad\quad a_1\leq z < a_1+a_2,
    \end{cases}
\end{equation}
where $a_1=0.2$, $a_2=0.8$, $w_1=0.3$, $w_2=0.07$, $u_1=0.1$, $u_2=0.01$ and $u_3=0.001$. Additionally, all the couplings to the $z$ axis are 0, i.e.
\begin{equation}
    \epsilon_{xz}=\epsilon_{zx}=\epsilon_{yz}=\epsilon_{zy}=0,
\end{equation}
and $\epsilon_{zz}=1$. The spectrum given by the corresponding transfer matrix is shown in Fig.~\ref{fig: EPS in PC}. Notice that the numerical values used in this example serve only as one convenient realisation. The in-plane permittivities chosen here are smaller than unity and can be understood as the effective response of an anisotropic metamaterial rather than that of a conventional passive dielectric. However, nothing in the argument depends on this choice, and the same phenomenology occurs for $\tilde{\epsilon}>\mathbb{1}_2$ at correspondingly rescaled frequencies. 

Related studies of anisotropic photonic crystals have investigated special band-edge degeneracies and the associated transmission resonances \cite{figotin2005gigantic}, while reciprocity has been used to constrain the transfer matrix structure in periodic photonic systems, and consequently, the pairing of Bloch wavevectors \cite{nada2017theory}. 

\section{Summary and discussion}

In this work, we have shown that the spectral gaps of Hermitian systems can themselves host robust structures when viewed through the transfer-matrix formalism. While conventional band topology focuses on Bloch states with real crystal momentum, states inside a gap are naturally associated with complex momenta and therefore probe a non-Hermitian extension of the band structure. From this perspective, the gap is not a featureless region separating bulk bands, but possesses its own spectral organisation encoded in the transfer matrix. We have demonstrated that this organisation can undergo a topological transition characterised by the creation or annihilation of exceptional points within the gap.

A central result of our work is that spinless TRS imposes strong constraints on the complex-momentum spectrum, allowing a $\mathbb{Z}_2$ index to be assigned to every energy inside a gap away from the exceptional points. In-gap exceptional points are the domain walls of this index, which is why an isolated one is stable against small symmetry-preserving perturbations and why, away from the band edges, they can only be created or annihilated in pairs. The transition between gaps with and without in-gap exceptional points is mediated by a crossing of the imaginary-momentum branches, in close analogy with the pair creation of Weyl points in three dimensions: what is protected is each exceptional point individually, rather than the existence of the energy window that they bound. The transition is therefore not associated with a change in the propagating bulk bands, which remain gapped throughout, but with a change in the structure of the evanescent spectrum inside the gap. 

Our results also clarify the physical significance of these in-gap exceptional points. For a bound state whose energy lies inside a gap and sufficiently close to a band edge, one might analytically predict its asymptotic behaviour to inherit the momentum of the nearest bulk extremum, leading to a purely exponential decay when the corresponding band edge occurs at zero crystal momentum. We have shown that this approximation is incomplete. When the energy of a bound state lies between a pair of in-gap exceptional points arising from the slowest-decaying modes, the relevant evanescent modes acquire a nonzero real momentum component despite the fact that the nearest band edge remains at zero crystal momentum. As a consequence, the tail of the bound state becomes oscillatory. The appearance of oscillations is therefore controlled not by the momentum of the band edge, but by the structure of the complex-momentum spectrum within the gap. In this sense, in-gap exceptional points provide a bulk characterisation of the asymptotic structure of localised states.

Although we introduced in-gap exceptional points here in the context of tight-binding models, the mechanism responsible for their existence is considerably more general. The spectral structure described in Sec.~\ref{sec: symmetry} follows from the interplay of TRS and a symplectic constraint on the transfer matrix. The origin of the latter depends on the setting: in classical systems, it is a generic property of transfer matrices associated with a Hamiltonian structure, whereas for quantum systems, it comes from the Hermiticity of the Hamiltonian operator. These ingredients are common to a broad range of both classical and quantum systems. We therefore expect the qualitative phenomenology described here—including the emergence of in-gap exceptional points and the associated transition in the evanescent spectrum—to extend well beyond the particular realisations considered in this work.

Several directions naturally follow from our results. First, while we focused on 1D systems with spinless TRS, it remains to understand how these ideas generalise to other symmetry classes and to higher-dimensional systems, where exceptional points may evolve into more intricate non-Hermitian nodal features. Second, the counting argument of Appendix~\ref{appendix: local topological index} shows that in-gap exceptional points come in pairs only when the index takes the same value just inside both band edges; whether a gap with an odd number of in-gap exceptional points can be realised, and what would then happen at its edges, is an open question. Third, the connection established in Ref.~\cite{saha2025arbitrary} between transfer-matrix exceptional points of higher order and higher-order van Hove singularities suggests that higher-order in-gap coalescences, which our $\mathbb{Z}_2$ index does not resolve, deserve a separate treatment.

More broadly, our work suggests that transfer matrices provide a natural language for extending topological band theory beyond propagating states. In the same way that Bloch bands reveal topological properties of the bulk spectrum, the complex-momentum spectrum reveals intrinsic spectral properties of the gaps. 

\begin{acknowledgments}
We thank Sébastien Guenneau, Marie Touboul, Andrés Perez Fadon and Henry Davenport for insightful discussions. This work was supported by a UKRI Future Leaders Fellowship MR/Y017331/1. This work was supported by the National Agency for Research and Development (ANID) / Scholarship Program/DOCTORADO BECAS CHILE/72230283. We also acknowledge the use of Claude by Anthropic and ChatGPT by OpenAI for assisting with derivations. This work was supported by the Royal Society via grants URF\textbackslash R\textbackslash221004 and RGF\textbackslash{}EA\textbackslash{}180121 and by the Engineering and Physical Sciences Research Council [grant numbers EP/W016141/1, EP/Y01510X/1 and UKRI2226]. 
\end{acknowledgments}

\newpage
\bibliography{References}
\newpage
\appendix

\section{Construction of a general transfer matrix for tight-binding Hamiltonians}
\label{appendix: construction general transfer matrix}
Here we provide a review of the construction of the transfer matrix for a general 1D translationally invariant tight-binding Hamiltonian, given that in the supercell representation, the matrix $J$ is singular. This corresponds to the process shown in Ref.~\cite{DwivediChua2016}. We start by performing a reduced singular value decomposition (SVD) $J=U\Sigma W^{\dagger}$ with $U,W$ being $\mathcal{N}\times r$ matrices and $\Sigma$ is an $r\times r$ matrix composed of the singular values of $J$. Here, $\mathcal{N}$ is the size of the chosen supercell and $r$ is the rank of $J$. Here, we also assume that $J^2=0$, which can always be ensured by the choice of a sufficiently large supercell and corresponds to the case $\mathcal{N}\geq 2qR$, that is, a supercell containing at least $2R$ unit cells. Note that the minimal grouping into supercells of $R$ unit cells, used in Sec.~\ref{sec: transfer matrix} to bring the model to nearest-neighbour form, need not satisfy $J^2=0$; doubling the supercell always does. Hence, we get the following relations between the matrices $U$ and $W$:
\begin{equation}
    U^{\dagger}U=W^{\dagger}W=\mathbb{1},\quad U^{\dagger}W=0.
\end{equation}
Now, we use the columns of $U$ and $W$ to write the supercell wavefunction at position $n$ as
\begin{equation}
    \Psi_n=U\alpha_{n}+W\beta_{n}+X\zeta_{n},
\end{equation}
where $\alpha_{n}$, $\beta_{n}$ and $\zeta_{n}$ are vector coefficients, while $X$ is a matrix constructed from vectors orthogonal to the columns of $U$ and $W$
\begin{equation}
    U^{\dagger}X= W^{\dagger}X=0,\quad X^{\dagger}X=\mathbb{1}.
\end{equation}
The recursion relation given by the Schrödinger equation in the bulk is
\begin{equation}
    J^{\dagger}\Psi_{n-1}+M\Psi_{n}+J\Psi_{n+1}=E\Psi_{n},
\end{equation}
which, for energies $E$ that are not eigenvalues of the onsite matrix $M$, can be written as
\begin{equation}
    \Psi_{n}=(E-M)^{-1}J\Psi_{n+1}+(E-M)^{-1}J^{\dagger}\Psi_{n-1}.
\end{equation}
The transfer matrix constructed below is therefore defined for $E\notin\mathrm{spec}\,M$; the isolated energies at which $E-M$ is singular can be treated by continuity, or removed by a different choice of supercell.
Using the orthogonality relations shown above, we can multiply from the left by $U^{\dagger}$, $W^{\dagger}$ and $X^{\dagger}$ to get the coefficients $\alpha_{n}$, $\beta_{n}$ and $\zeta_{n}$ respectively on the left hand side of the equation. Explicitly, 
\begin{equation}
    \begin{aligned}
        \alpha_{n}=&G_{UU}\Sigma \beta_{n+1}+G_{WU}\Sigma\alpha_{n-1}\\
        \beta_{n}=&G_{UW}\Sigma \beta_{n+1}+G_{WW}\Sigma\alpha_{n-1}\\
        \zeta_{n}=&G_{UX}\Sigma \beta_{n+1}+G_{WX}\Sigma\alpha_{n-1},
    \end{aligned}
\end{equation}
where we have defined 
\begin{equation}
    G_{AB}:=B^{\dagger}(E-M)^{-1}A,\quad A,B\in\{U,W,X\}.
\end{equation}
Then, this allows us to express, for instance, $\beta_{n+1}$ in terms of $\beta_{n}$ and $\alpha_{n-1}$. Therefore, we can find a matrix $T$ such that
\begin{equation}
    \begin{pmatrix}
        \beta_{n+1}\\\alpha_{n}
    \end{pmatrix}=T\begin{pmatrix}
        \beta_{n}\\\alpha_{n-1}
    \end{pmatrix}.
\end{equation}
This matrix $T$ is the transfer matrix and it is given by
\begin{equation}
    T=\begin{pmatrix}
        \Sigma^{-1}G_{UW}^{-1} & -\Sigma^{-1}G_{UW}^{-1}G_{WW}^{\phantom -}\Sigma\\
        G_{UU}^{\phantom -}G_{UW}^{-1} & (G_{WU}^{\phantom -}-G_{UU}^{\phantom -}G_{UW}^{-1}G_{WW}^{\phantom -})\Sigma
    \end{pmatrix},
\end{equation}
which is a function of $E$ given by the $G$ matrices.

\section{Properties of the transfer matrix and existence of exceptional points}
\label{appendix: Existence of EP}
Spinless TRS means that the Hamiltonian $H$ is invariant under complex conjugation. That is, if we represent the complex conjugation operator by $\mathcal{K}$, then $\mathcal{K}H\mathcal{K}^{-1}=H$ with $\mathcal{K}^2=\mathbb{1}$. This implies the following condition on the Bloch Hamiltonian for real crystal momentum:
\begin{equation}
    \begin{aligned}
        H_{\mathrm{B}}(k)&=H^*_{\mathrm{B}}(-k).
    \end{aligned}
\end{equation}
Let us consider first the transfer matrix $\hat{T}(E)$ of the Dirac Hamiltonian $H(k)=\gamma k+m$ for certain $\gamma$ and $m$
\begin{equation}
    \hat{T}(E):=e^{\mathrm{i}\gamma^{-1}(E-m)}.
\end{equation}
 Applying the spinless TRS condition to the Dirac Hamiltonian results in the conditions $\gamma^*=-\gamma$ and $m^*=m$. Consequently, the transfer matrix must be a real matrix: $\mathcal{K}\hat{T}(E)\mathcal{K}=\hat{T}(E)$ where $\mathcal{K}$ is the conjugate operator.

For the general case of the transfer matrix we start by considering the recursion relation for the supercells
\begin{equation}
    J^\dagger\Psi_{n-1}+M\Psi_n+J\Psi_{n+1}=E\Psi_n,
\end{equation}
with 
\begin{equation}
    \Psi_{n}:=\begin{pmatrix}
        \psi_{(n-1)R+1}\\
        \psi_{(n-1)R+2}\\
        \vdots\\
        \psi_{nR}
    \end{pmatrix}\in\mathbb{C}^{qR},
\end{equation}
and $M$ being the onsite matrix and $J$ the hopping matrix connecting supercells, following the notation of \cite{DwivediChua2016}. Where the construction of Appendix~\ref{appendix: construction general transfer matrix} is used, the supercell is understood to be enlarged so that $J^2=0$. Also, $R$ is the range of hopping and $q$ is the number of orbitals per unit cell. As this is just a relabelling of the lattice sites, we can write our Hamiltonian as
\begin{equation}
    H=\sum_{\tilde{r},\tilde{\alpha},\tilde{\beta}} M \ket{\tilde{r},\tilde{\alpha}}\bra{\tilde{r},\tilde{\beta}}+J\ket{\tilde{r}+1,\tilde{\alpha}}\bra{\tilde{r},\tilde{\beta}}+\mathrm{h.c.},
\end{equation}
where the sum runs over the new indices of the new lattice. Now, since we have only nearest-neighbour hopping, we write the Bloch Hamiltonian as
\begin{equation}
    H_{\mathrm{B}}(k)=M+Je^{\mathrm{i}k}+J^\dagger e^{-\mathrm{i}k}.
\end{equation}
Thus, if $H_{\mathrm{B}}$ obeys spinless TRS, we conclude that
\begin{equation}
    \begin{aligned}
        M^*&=M,\\
        J^*&=J.
    \end{aligned}
\end{equation}
In order to obtain the general form of the transfer matrix, we perform a reduced singular value decomposition (SVD) of $J$
\begin{equation}
    J=Z\Sigma V^\dagger,
\end{equation}
where the columns of $V$ are $r$ complex vectors of dimension $\mathcal{N}$ that form an orthonormal basis for the row space of $J$. For the general nonreduced SVD, the columns of $V$ include an orthonormal basis of the whole space since the remaining $\mathcal{N}-r$ column vectors are an orthonormal basis of the null space of $J$. Here, $r$ is the rank of $J$. The columns of $Z$ are $r$ complex vectors of dimension $\mathcal{N}$ that form an orthonormal basis for the column space of $J$ and $\Sigma$ is an $ r \times r$ diagonal matrix with diagonal elements being real positive numbers (called the singular values of $J$). \\
If we restrict $J$ to be a real matrix, then it is always possible to construct a real SVD $J=U\Sigma W^T$. The matrices are related according to the following: let us consider first a general nonreduced SVD $J=Z\Sigma V^\dagger$ and the real nonreduced SVD $J=U\Sigma W^T$. Then we study the following equivalent matrices
\begin{equation}
    \begin{aligned}
        J^\dagger J&=V\Sigma^2 V^\dagger,\\
        J^T J &= W\Sigma^2 W^T.
    \end{aligned}
\end{equation}
Let $L:=J^\dagger J$, a real square matrix. Both right-hand sides correspond to the diagonal decomposition of the matrix $L$. Thus, the column vectors of $W$ and $V$ are a set of eigenvectors of the matrix $L$. Let us analyse first the case where the diagonal entries of $\Sigma$ do not repeat. In that case we have
\begin{equation}
    \begin{aligned}
        Lw_i&=\sigma^2_i w_i,\\
        Lv_i&=\sigma^2_i v_i
    \end{aligned}
\end{equation}
which means that $v_i=e^{\mathrm{i}\theta_i}w_i$ since both vectors have norm 1. Thus, in that case
\begin{equation}
    V=WP,
\end{equation}
where $P=\mathrm{diag}(e^{\mathrm{i}\theta_1},\ldots,e^{\mathrm{i}\theta_n})$. In the case that we have repeated eigenvalues, now the vectors $v_i$ can be written as a linear combination of the basis vectors associated with the eigenspace of the repeated eigenvalue. Thus
\begin{equation}
    V=WP,
\end{equation}
where $P$ now is a matrix that has diagonal elements $e^{\mathrm{i}\theta_i}$ when the eigenvalue is not repeated and it is block-diagonal for those entries that have repeated eigenvalues. This matrix $P$ must be unitary since
\begin{equation}
    \begin{aligned}
        V^\dagger V=&(WP)^\dagger WP\\
        =&P^\dagger P\\
        \implies &P^\dagger P=\mathbb{1}.
    \end{aligned}
\end{equation}
We can perform the same analysis for the matrix $\tilde{L}:=JJ^\dagger$, from which we can conclude that $Z=UQ$ with a unitary $Q$. Now, since the $r$ singular values are positive real numbers, the associated vectors are not linear combinations of the vectors associated with the null space. Therefore, we can write: $V_r=W_rP_r$ and $Z_r=U_rQ_r$, where the subscript $r$ means the associated matrix with the reduced SVD. We can show now that $Q_r=P_r$ blockwise
\begin{equation}
    \begin{aligned}
        J=U_r\Sigma_r W_r^T&= Z_r\Sigma_r V_r^\dagger\\
        &=U_r (Q_r\Sigma_r P_r^\dagger)W^T_r\\
        \implies \Sigma_r&=(Q_r\Sigma_r P_r^\dagger)\\
        \implies P_r&=Q_r.
    \end{aligned}
\end{equation}
where this equality holds for every block corresponding to the respective singular value, and we also used that $\Sigma_r$ is invertible (as opposed to $\Sigma$).
So for a general SVD $J=Z_r\Sigma_r V^\dagger_r$
\begin{equation}
    \Psi_n=Z_r\alpha_n+V_r\beta_n+X\zeta_n,
\end{equation}
where the term $X\zeta_n$ accounts for the dimensions not covered by the columns of $Z_r$ and $V_r$: $Z^\dagger_r X=V^\dagger_r X=0$. Therefore,
\begin{equation}
\label{eq: eqs for beta alpha}
    \begin{aligned}
        J\Psi_n &=Z_r\Sigma_r\beta_n,\\
        J^\dagger\Psi_n &=V_r\Sigma_r\alpha_n,
    \end{aligned}
\end{equation}
from which we arrive at the transfer matrix equation
\begin{equation}
    \begin{pmatrix}
        \beta_{n+1}\\
        \alpha_{n}
    \end{pmatrix}=T(E)\begin{pmatrix}
        \beta_{n}\\
        \alpha_{n-1}
    \end{pmatrix},
\end{equation}
where the matrix $T(E)$ is composed of the matrices $Z_r,V_r$ and $G(E)$. Then, we insert the transformations in Eq.~\eqref{eq: eqs for beta alpha}:
\begin{equation}
    \begin{aligned}
        J\Psi_n &=U_rP_r\Sigma_r\beta_n\\
        &=U_r\Sigma_r (P_r\beta_n)\\
        &=U_r\Sigma_r \tilde{\beta}_n,\\
        J^\dagger\Psi_n &=W_rP_r\Sigma_r\alpha_n\\
        &=W_r\Sigma_r(P_r\alpha_n)\\
        &=W_r\Sigma_r\tilde{\alpha}_n,
    \end{aligned}
\end{equation}
where we named $\tilde{\beta}_n:=P_r\beta_n$ and $\tilde{\alpha}_n:=P_r\alpha_n$. Following the exact same procedure to get the transfer matrix we obtain
\begin{equation}
    \begin{pmatrix}
        \tilde{\beta}_{n+1}\\
        \tilde{\alpha}_{n}
    \end{pmatrix}=\tilde{T}(E)\begin{pmatrix}
        \tilde{\beta}_{n}\\
        \tilde{\alpha}_{n-1}
    \end{pmatrix},
\end{equation}
with $\tilde{T}(E)$ being a purely real matrix. Calling $\Phi_n:=(\beta_{n}, \alpha_{n-1}) $ and similarly for $\tilde{\Phi}_n$, we transform back to the original variables
    \begin{equation}
    \mathcal{P}\Phi_{n+1}=\tilde{T}(E)\mathcal{P}\Phi_n,
\end{equation}
where 
\begin{equation}
    \mathcal{P}:=\begin{pmatrix}
        P_r &0\\0&P_r
    \end{pmatrix}.
\end{equation}
Then we conclude that
\begin{equation}
    \mathcal{P}^{-1}\tilde{T}(E)\mathcal{P}=T(E),
\end{equation}
which means that the transfer matrix $T(E)$ is always similar to the real transfer matrix $\tilde{T}(E)$. This implies that their spectrum is the same and consequently, the eigenvalues of the matrix either are purely real or come in conjugated pairs of complex numbers. Thus the condition becomes
\begin{equation}
\tilde{\mathcal{V}} T(E) \tilde{\mathcal{V}}=T(E),\quad \tilde{\mathcal{V}}^2=\mathbb{1}_{2r\times 2r},
\end{equation}
where $\tilde{\mathcal{V}}:=\mathcal{P}^\dagger \mathcal{K}\mathcal{P}$. In particular, by choosing a real SVD, one obtains 
\begin{equation}
    \mathcal{K} T(E) \mathcal{K}=T(E),\quad \mathcal{K}^2=1.
\end{equation}
\subsection{Symplectic condition}
To derive the symplectic condition of the transfer matrix for the 1D lattice, we work in the supercell basis and consider two solutions $\psi_n$ and $\phi_n$ of the bulk recursion relation
\begin{equation}
    M\psi_{n}+J\psi_{n+1}+J^{\dagger}\psi_{n-1}=E\psi_{n},
\end{equation}
and likewise for $\phi_n$, at the same real energy $E$. These solutions need not be normalisable eigenstates of $H$; for $E$ inside a gap they are generically evanescent. The global relation $\braket{H\phi|\psi}=\braket{\phi|H\psi}$ therefore cannot be used directly. Instead, we use a local identity. Multiplying the recursion for $\psi$ at supercell $n$ from the left by $\phi_n^{\dagger}$, and the Hermitian conjugate of the recursion for $\phi$ at supercell $n$ from the right by $\psi_n$, both sides become $E\,\phi_n^{\dagger}\psi_n$, where we used $M^{\dagger}=M$ and $E\in\mathbb{R}$. Hence
\begin{equation}
 \phi^{\dagger}_{n}(M\psi_{n}+J\psi_{n+1}+J^{\dagger}\psi_{n-1})=(\phi^{\dagger}_{n}M+\phi^{\dagger}_{n+1}J^{\dagger}+\phi^{\dagger}_{n-1}J)\psi_{n}.
\end{equation}
This identity uses only the Hermitian structure of the hoppings and the reality of $E$. By rearranging it, we get
\begin{equation}
\begin{aligned}
\label{eq: Wronskian equality}
    -\phi^{\dagger}_{n}J\psi_{n+1}+\phi^{\dagger}_{n+1}J^{\dagger}\psi_{n}&=-\phi^{\dagger}_{n-1}J\psi_{n}+\phi^{\dagger}_{n}J^{\dagger}\psi_{n-1}\\
    \mathcal{W}_{n}(\phi,\psi)&=\mathcal{W}_{n-1}(\phi,\psi).
    \end{aligned}
\end{equation}
The quantity $\mathcal{W}_{n}(\phi,\psi)$ is called the Wronskian and according to the equation above, it is invariant under changes in the index $n$. We write the Wronskian in vector notation as
\begin{equation}
    \begin{aligned}
        \mathcal{W}_{n}(\phi,\psi)=&\chi_n(\phi)^{\dagger}\mathcal{V}\chi_n(\psi),\\
        \chi_n(f):=&\begin{pmatrix}
            f_{n+1}\\f_{n}
        \end{pmatrix},\\
        \mathcal{V}:=&\begin{pmatrix}
            0& J^{\dagger}\\
            -J&0
        \end{pmatrix}.
    \end{aligned}
\end{equation}
By construction, we know that the transfer matrix $T$ propagates the coefficients, so we write
\begin{equation}
    T\chi_{n-1}(f)=\chi_{n}(f).
\end{equation}
Plugging this into Eq.~\eqref{eq: Wronskian equality}, we get
\begin{equation}
    \begin{aligned}
        \chi_{n-1}(\phi)^{\dagger}T^{\dagger}\mathcal{V}T\chi_{n-1}(\psi)&=\chi_{n-1}(\phi)^{\dagger}\mathcal{V}\chi_{n-1}(\psi)\\
    \end{aligned}
\end{equation}
since this must be valid for any $n$ and any pair of solutions, whose initial data $\chi_{n-1}$ can be chosen freely, we conclude that
\begin{equation}
\label{eq: symplectic cond easy T}
    T^{\dagger}\mathcal{V}T=\mathcal{V}.
\end{equation}
Notice also that $\mathcal{W}_{n}(\psi,\psi)=\mathrm{i}\hbar\,j_{n,n+1}$, where $j_{n,n+1}$ is the probability current between sites $n$ and $n+1$. So if this quantity does not depend on $n$ and we know that
\begin{equation}
    \frac{\mathrm{d}\rho_n}{\mathrm{d}t}=\frac{\mathrm{i}}{\hbar}\left[\mathcal{W}_{n}(\psi,\psi)-\mathcal{W}_{n-1}(\psi,\psi)\right]=j_{n-1,n}-j_{n,n+1},
\end{equation}
where $\rho_n:=\psi^{\dagger}_n\psi_n$, then it follows that
\begin{equation}
    \frac{\mathrm{d}\rho_n}{\mathrm{d}t}=0.
\end{equation}
Notice that Eq.~\eqref{eq: symplectic cond easy T} is for a transfer matrix that propagates the coefficients given by the full vector $\Psi_n$. However, the transfer matrix used in this paper propagates the left and right singular subspaces of the hopping matrix $J$. That is, the degrees of freedom associated with the column and row spaces. Nevertheless, the procedure to find the symplectic condition for this matrix is similar.
Given the SVD of $J$ we expand the components of the two solutions $\psi_n$ and $\phi_n$ introduced above as
\begin{equation}
    \begin{aligned}
        \psi_{n}&=U\alpha_{n}+W\beta_{n}+X\zeta_{n},\\
        \phi_{n}&=U\tilde{\alpha}_{n}+W\tilde{\beta}_{n}+X\tilde{\zeta}_{n},
    \end{aligned}
\end{equation}
where the tilde denotes the coefficients of $\phi$. Then, the Wronskian is
\begin{equation}
    \begin{aligned}
        \mathcal{W}_{n}(\phi,\psi)&=-\tilde{\alpha}_{n}^{\dagger}\Sigma\beta_{n+1}+\tilde{\beta}_{n+1}^{\dagger}\Sigma\alpha_{n}\\
        &=\tilde{\eta}_{n}^{\dagger}\mathcal{F}\eta_{n}
    \end{aligned}
\end{equation}
where
\begin{equation}
    \mathcal{F}:=\begin{pmatrix}
        0&\Sigma\\
        -\Sigma & 0
    \end{pmatrix}.
\end{equation}
Also, $\eta_{n}:=(\beta_{n+1},\alpha_{n})$ and similarly for $\tilde{\eta}$. Using again that the Wronskian is invariant under changes in $n$ and that the transfer matrix propagates these degrees of freedom, i.e.
\begin{equation}
    T(E)\eta_{n}=\eta_{n+1},
\end{equation}
we obtain the condition
\begin{equation}
    T(E)^{\dagger}\mathcal{F}T(E)=\mathcal{F}.
\end{equation}

\subsection{Exceptional points and defectiveness of the transfer matrix}
\label{app: defectiveness of T}
Using these properties, we study first the action of the conjugate operator on the eigenstates of the transfer matrix
\begin{equation}
\label{eq: transfer matrix eigenvectors eigenspaces}
    \begin{aligned}
        \hat{T}(E)\psi&=z \psi\\
        \mathcal{K}\hat{T}(E)\psi&=z^* \mathcal{K}\psi\\
        \hat{T}(E)\mathcal{K}\psi&=z^* \mathcal{K}\psi.
    \end{aligned}
\end{equation}
Thus, if $z\notin\mathbb{R}$ and the eigenspace of $z$ is nondegenerate, $\mathcal{K}\psi$ is another eigenvector of $\hat{T}(E)$ with eigenvalue $z^*$. On the other hand, if $z\in\mathbb{R}$ and the eigenspace of $z$ is nondegenerate, then $\mathcal{K}\psi=\alpha \psi$ for some $\alpha\in\mathbb{C}$. Regarding this $\alpha$ we know that
\begin{equation}
    \begin{aligned}
        \mathcal{K}\psi&=\alpha \psi\\
        \mathcal{K}^2\psi&=\alpha^*\mathcal{K}\psi\\
        \psi&=|\alpha|^2\psi,
    \end{aligned}
\end{equation}
which implies that $|\alpha|^2=1$ and hence, $\alpha$ can be written as a phase factor $\alpha=e^{\mathrm{i}\theta}$, that we can get rid of by the gauge transformation $\psi\rightarrow e^{\mathrm{i}\theta/2}\psi$.

On its own, the complex symplectic condition is a similarity transformation between $T(E)^{\dagger}$ and $T(E)^{-1}$ given by
\begin{equation}
\label{eq_app: complex symplectic condition}
    T(E)^{\dagger}=\mathcal{F}T(E)^{-1}\mathcal{F}^{-1}.
\end{equation}
This means that if $z$ is an eigenvalue of $T(E)$, then $1/z^*$ is also an eigenvalue of $T(E)$. Therefore, there is a transformation $\mathcal{X}$ that sends right eigenvectors in the eigenspace of $z$ to the left eigenspace of $1/z^*$. This can be seen as
\begin{equation}
    T\phi=z\phi\implies \phi^{\dagger}T^{\dagger}=z^*\phi^{\dagger}\implies (\phi^{\dagger}\mathcal{F})T=\frac{1}{z^*}(\phi^{\dagger}\mathcal{F}),
\end{equation}
where we plugged in Eq.~\eqref{eq_app: complex symplectic condition}. More precisely, we define $\mathcal{X}$ as
\begin{equation}
    \begin{aligned}
        \mathcal{X}:R_{z}&\rightarrow L_{1/z^*},\\
        \phi&\mapsto \phi^{\dagger}\mathcal{F}.
    \end{aligned}
\end{equation}
where $R_{z}$ and $L_{z}$ denote the right and left eigenspace of the eigenvalue $z$. The map $\mathcal{X}$ is antilinear, continuous, and injective because $\mathcal{F}$ is invertible; in particular, $\mathcal{X}(\phi)\neq 0$ whenever $\phi\neq 0$.

We now distinguish two kinds of collisions. At band edges, real allowed momenta merge on the real axis and continue as evanescent momenta. Inside the gap, imaginary allowed momenta merge on the imaginary axis and split into the complex plane.

We first consider a band edge. The argument is the same at both edges of the gap, so we consider one of them, at energy $E_{\mathrm{EP}}$, and assume that the band lies below it. Let $z_{1}(E)$ and $z_{2}(E)$ be the two eigenvalues that merge at $E_{\mathrm{EP}}$, with $z_{1}(E_{\mathrm{EP}})=z_{2}(E_{\mathrm{EP}})$, and assume that for $E\neq E_{\mathrm{EP}}$ close to $E_{\mathrm{EP}}$ the spectrum is nondegenerate, so that all right and left eigenspaces are one-dimensional.

For $E<E_{\mathrm{EP}}$ the two momenta are real, so $|z_{i}|=1$, i.e.\ $1/z_{i}^{*}=z_{i}$: each eigenvalue is its own partner, and $\mathcal{X}$ maps $R_{z_{i}}$ into $L_{z_{i}}$. For $E>E_{\mathrm{EP}}$ the momenta are purely imaginary, so $|z_{1}|\neq 1$ and $1/z_{1}^{*}\neq z_{1}$. Hence $1/z_{1}^{*}=z_{2}$: above the band edge the two merging eigenvalues are each other's partners, and $\mathcal{X}$ maps $R_{z_{1}}$ into $L_{z_{2}}$ and $R_{z_{2}}$ into $L_{z_{1}}$. Since these eigenspaces are one-dimensional,
\begin{equation}
\begin{aligned}
    \mathcal{X}\big(\phi_{1}(E)\big)\propto
    \begin{cases}
        \chi_{1}(E), & E<E_{\mathrm{EP}},\\
        \chi_{2}(E), & E>E_{\mathrm{EP}}.
    \end{cases}
\end{aligned}
\end{equation}
Thus, since $\phi_{1}(E)$ can be chosen a continuous function, at the splitting point the conjugate operator $\mathcal{X}$ has to map the eigenvector $\phi_{1}(E_{\mathrm{EP}})$ to $\chi_{1}(E_\mathrm{EP})$ and to $\chi_{2}(E_\mathrm{EP})$. This implies that at this point, $\chi_{1}(E_\mathrm{EP})\propto\chi_{2}(E_\mathrm{EP})$. Finally, this implies the coalescence of the eigenvectors of the transfer matrix
\begin{equation}
    \phi_{1}(E_\mathrm{EP})^{\dagger}F\propto \phi_{2}(E_\mathrm{EP})^{\dagger}F\implies \phi_{1}(E_\mathrm{EP})\propto \phi_{2}(E_\mathrm{EP}).
\end{equation}

\begin{figure}
    \centering
    \includegraphics[width=0.95\linewidth]{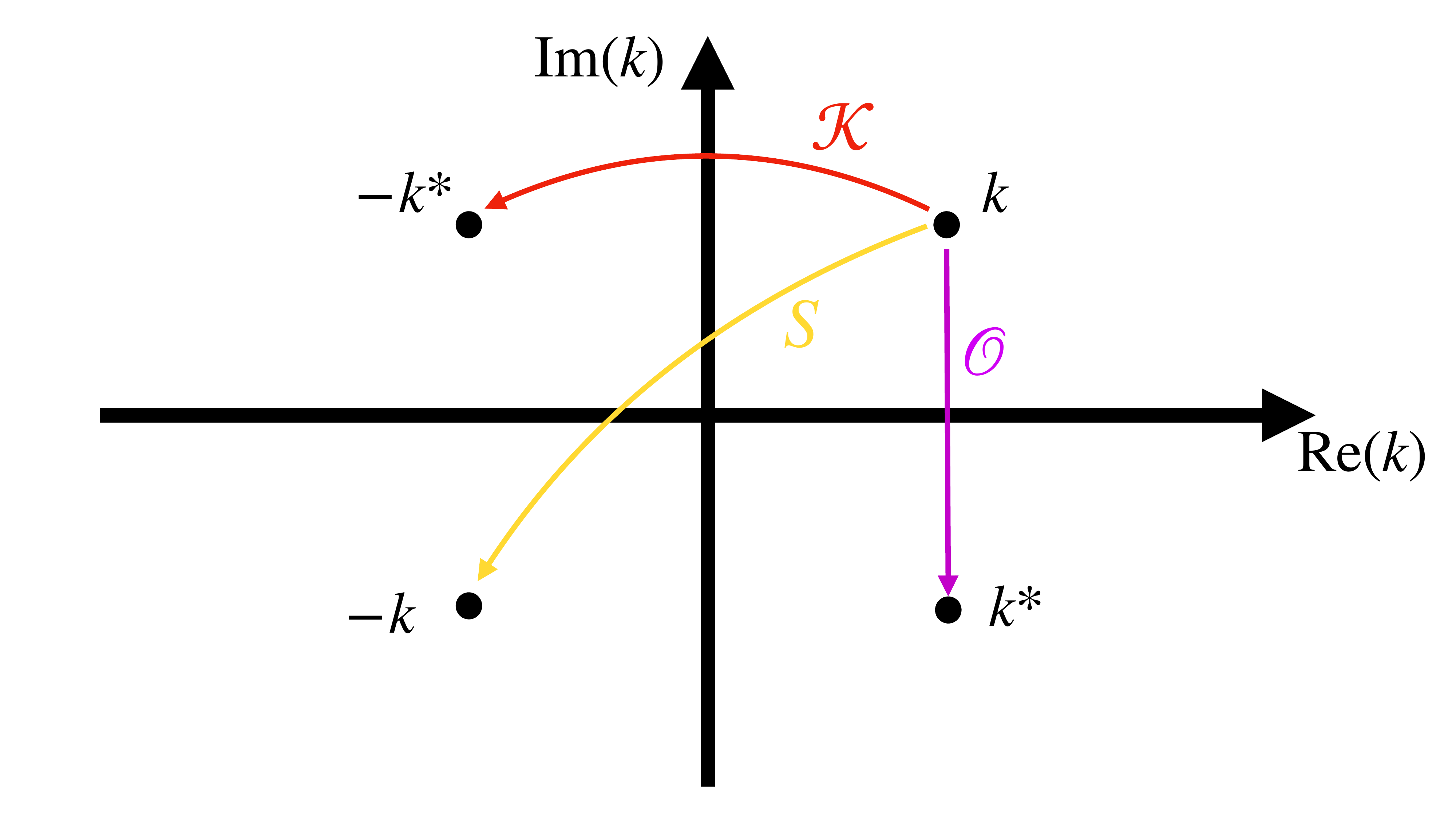}
    \caption{Action of the antiunitary operator $\mathcal{K}$, the intertwiner $S$ and their composition $\mathcal{O}=S\mathcal{K}$ on the transfer-matrix eigenstate with momentum $k$: they map it to the eigenstates with momenta $-k^*$, $-k$ and $k^*$, respectively.}
    \label{fig:KSP map}
\end{figure}

If we include now TRS, we can map the eigenstate associated with momentum $k$ to the eigenstates corresponding to $-k^*,-k$ and $k^*$ via the operators $\mathcal{K},S$ and $\mathcal{O}$, where $S$ intertwines the transfer matrix with its inverse, $T(E)=ST(E)^{-1}S^{-1}$, so that it maps the eigenstate with momentum $k$ to the one with momentum $-k$, and $\mathcal{O}=S\mathcal{K}$ as depicted in Fig.~\ref{fig:KSP map}. Thus, we can study the case where imaginary (real) momenta merge on the imaginary (real) axis and then split into the complex plane. The central idea is that in both cases, there is always an operator that maps an eigenspace to itself at some energies and to another eigenspace for other energies. Hence, at the interface between these two energy regions (the splitting point), those states have to be the same. For the splitting of the imaginary momenta, that operator is $\mathcal{K}$ whereas for the splitting of real momenta the operator is $\mathcal{O}$.

So far, we have shown that the eigendirections associated with two symmetry-related momentum branches coalesce when the branches meet and leave a symmetry-invariant axis of the complex-momentum plane. However, this fact alone does not exclude the possibility that the repeated transfer-matrix eigenvalue possesses a second, linearly independent eigenvector at the collision. We now show that this possibility is excluded for the branch reconnections considered here.

We first consider an in-gap collision occurring at
\begin{equation}
    k_0=\mathrm{i}q_0,
    \qquad q_0\in\mathbb{R}.
\end{equation}
In a neighbourhood of the collision, we introduce a signed parameter $s$ and write one of the momentum branches as
\begin{equation}
    k(s)=s+\mathrm{i}q(s),
    \qquad k(0)=k_0,
\end{equation}
where $q$ is analytic with real Taylor coefficients at $s=0$. On the complex side of the collision, $s$ runs over a real interval, so that $\mathrm{Re}\,k=s$ and $\mathrm{Im}\,k=q(s)$. Spinless TRS associates this momentum with
\begin{equation}
    -k(s)^*=-s+\mathrm{i}q(s),
\end{equation}
at the same energy. We choose the signed parametrisation such that
\begin{equation}
    k(-s)=-k(s)^*.
\end{equation}
It follows that
\begin{equation}
    q(-s)=q(s),
    \qquad
    E(-s)=E(s).
\end{equation}
Therefore, the energy is an even function of $s$, and its local expansion takes the form
\begin{equation}
    E(s)=E_0+E_2s^2+E_4s^4+\cdots,
\end{equation}
which implies
\begin{equation}
    E'(0)=0.
\end{equation}
A collision on the line $\mathrm{Re}\,k=\pi$, at $k_0=\pi+\mathrm{i}q_0$, is treated identically with $k(s)=\pi+s+\mathrm{i}q(s)$, since $-k^*\equiv k$ modulo $2\pi$ on that line.

Let $\phi(s)$ be a differentiable eigenvector branch, satisfying
\begin{equation}
    T(E(s))\phi(s)=z(s)\phi(s),
    \qquad
    z(s)=e^{\mathrm{i}k(s)}.
\end{equation}
Differentiating with respect to $s$ gives
\begin{equation}
\begin{aligned}
    T'(E(s))E'(s)\phi(s)
    +T(E(s))\phi'(s)
    = &z'(s)\phi(s)\\
    &+z(s)\phi'(s).
\end{aligned}
\end{equation}
Evaluating this expression at $s=0$, and using $E'(0)=0$, yields
\begin{equation}
    \left[T(E_0)-z_0\mathbb{1}\right]\phi'(0)
    =
    z'(0)\phi(0),
    \qquad
    z_0=e^{\mathrm{i}k_0}.
\end{equation}
Moreover, since $q(s)$ is even, $q'(0)=0$, and hence
\begin{equation}
    k'(0)=1,
    \qquad
    z'(0)\neq0.
\end{equation}
Defining
\begin{equation}
    A:=T(E_0)-z_0\mathbb{1},
    \qquad
    w_0:=\frac{\phi'(0)}{z'(0)},
\end{equation}
we obtain
\begin{equation}
    A\phi(0)=0,
    \qquad
    Aw_0=\phi(0).
\end{equation}
Thus, the nonzero vector $\phi(0)$ belongs simultaneously to the kernel and the image of $A$:
\begin{equation}
    0\neq\phi(0)\in\ker A\cap\mathrm{Im}A.
\end{equation}
If $T(E_0)$ were diagonalisable, then
\begin{equation}
    \ker A\cap\mathrm{Im}A=\{0\}.
\end{equation}
The nontrivial intersection obtained above therefore proves that $T(E_0)$ is defective. Equivalently, $\phi(0)$ and $w_0$ form a Jordan chain of length two:
\begin{equation}
    Aw_0=\phi(0),
    \qquad
    A\phi(0)=0.
\end{equation}
When only the two branches under consideration participate in the collision, the repeated eigenvalue has algebraic multiplicity two and geometric multiplicity one, and the collision is therefore a second-order exceptional point.

The argument above is not restricted to a collision on the imaginary-momentum axis. Its essential ingredients are that the two branches and their eigenvectors are analytic in a regular parameter $s$, that $E'(0)=0$, and that $k'(0)\neq0$. Above, $E'(0)=0$ followed from the symmetry relating the momenta at $s$ and $-s$ at the same energy.

For a collision in which real-momentum branches leave the real axis and become complex, one may write
\begin{equation}
    k_{\pm}(s)=q(s)\pm\mathrm{i}s,
\end{equation}
where the two branches are related by complex conjugation (Hermiticity). For a band-edge collision, at which a pair of real-momentum branches becomes a pair of imaginary-momentum branches, one may parametrise the two branches on the decaying side as

\begin{equation}
    k_{\pm}(s)=k_0\pm\mathrm{i}s,\quad k_0\in\{0,\pi\}.
\end{equation}

The momenta related by Hermiticity correspond to the same energy, so that $E(s)$ is even, while the chosen parametrisation ensures $k'(0)\neq0$. The proof of defectiveness is therefore identical to the one given above.

\subsection{$\mathbb{Z}_2$ index for the evanescent transfer spectrum}
\label{appendix: local topological index}

Here we construct the $\mathbb{Z}_2$ index $\nu(E)$ used in the main text. It is defined at every energy inside a bulk gap at which no two decaying multipliers coincide, and the generic in-gap exceptional points appear as the domain walls at which it jumps.

\subsubsection{The decaying spectrum is closed under complex conjugation}

Fix a real energy $E$ inside a bulk gap. No eigenvalue of $T(E)$ then lies on the unit circle, so the transfer-matrix spectrum splits unambiguously into the decaying multipliers
\begin{equation}
    \mathcal{Z}_{<}(E):=\left\{z\in\mathrm{spec}\,T(E)\;:\;|z|<1\right\}
\end{equation}
and the growing ones, with $|z|>1$. As shown above, spinless TRS allows $T(E)$ to be chosen real, so its spectrum is invariant under complex conjugation; since $|z^{*}|=|z|$, the subset $\mathcal{Z}_{<}(E)$ is itself invariant. The monic polynomial whose roots are the decaying multipliers,
\begin{equation}
    p_{E}(z):=\prod_{z_j\in\mathcal{Z}_{<}(E)}\left(z-z_j\right),
\end{equation}
therefore has real coefficients, and its discriminant
\begin{equation}
\label{appendix: def D(E)}
    \mathcal{D}(E):=\prod_{i<j}\left(z_i-z_j\right)^{2}
\end{equation}
is real. This is what allows a sign, and hence a $\mathbb{Z}_2$ label, to be attached to each in-gap energy. Both ingredients are needed. That $E$ lies in a gap keeps every multiplier off the unit circle, and Hermiticity, through $T(E)^{\dagger}\sim T(E)^{-1}$, organises them into reciprocal pairs $z,1/z^*$, so that $\mathcal{Z}_{<}(E)$ and its complement have equal size and the splitting into decaying and growing modes is unambiguous. Spinless TRS then makes $T(E)$ real, so that $\mathcal{Z}_{<}(E)$ is closed under complex conjugation and $p_E$ has real coefficients.

\subsubsection{Definition of the index}

At any in-gap energy at which the decaying multipliers are pairwise distinct we set
\begin{equation}
\label{appendix: def nu}
    \nu(E):=\frac{1-\mathrm{sgn}\,\mathcal{D}(E)}{2}\in\{0,1\}.
\end{equation}
For a real polynomial with simple roots one has $\mathrm{sgn}\,\mathcal{D}=(-1)^{c}$, where $c$ is the number of complex-conjugate root pairs. Hence
\begin{equation}
    \nu(E)=c(E)\ \bmod\ 2,
\end{equation}
with $c(E)$ the number of complex-conjugate pairs in $\mathcal{Z}_{<}(E)$. In particular, $\nu(E)=0$ whenever every decaying multiplier is real, i.e.\ every decaying mode has $\mathrm{Re}\,k\in\{0,\pi\}$ and decays purely exponentially, while $\nu(E)=1$ implies that at least one conjugate pair, of oscillatory decaying multipliers, i.e. $z$ and $z^*$, is present. The converse of the first statement holds for at most three decaying branches: with four or more decaying branches, $\nu=0$ is also compatible with an even number of conjugate pairs. With at most two decaying branches, as in all models of this paper, $\nu(E)=0$ and $\nu(E)=1$ correspond one-to-one to purely exponential and oscillatory bound-state tails (Sec.~\ref{sec: physical consequences} and Appendix~\ref{appendix: Asymptotic tails}). In general, the tail is set by the slowest-decaying mode, which $\nu$ alone does not identify.

Since $\mathcal{D}(E)$ is a continuous real function of $E$ which vanishes only when two decaying multipliers coincide, $\nu(E)$ is constant on each interval between consecutive zeros of $\mathcal{D}$. When $\mathcal{D}$ changes sign, two real decaying multipliers turn into a complex-conjugate pair, and the transfer matrix is defective at that energy (Appendix~\ref{app: defectiveness of T}); $\nu$ then changes by one. Every domain wall of $\nu$ is therefore an in-gap exceptional point, and generic in-gap exceptional points are precisely these domain walls. The converse is not true for every in-gap exceptional point: a fine-tuned exceptional point can occur at a zero of $\mathcal{D}$ at which $\nu$ does not change. An example is the transition point $|b|=|c|\neq0$ of the Dirac model of Eq.~\eqref{eq: Dirac model with EPs}: there the two decaying multipliers touch at $E=0$, where the transfer matrix is in fact defective, but they remain real on both sides, so that $\mathcal{D}$ vanishes at $E=0$ without changing sign. A small change of $c$ either removes this zero or splits it into a pair of sign-changing zeros, i.e., a pair of generic in-gap exceptional points. Thus $\nu$ tracks whether the decaying multipliers are real or complex, i.e., whether the corresponding evanescent waves decay purely exponentially or with oscillations; generic in-gap exceptional points are where this changes, whereas a fine-tuned one, such as the point $|b|=|c|$, is a degeneracy of $T(E)$ across which it does not.

Whether $T(E)$ is defective at such a fine-tuned zero is not fixed by the sign of $\mathcal{D}$. Restricting $T(E)$ to the two-dimensional invariant subspace of the colliding multipliers gives a $2\times2$ matrix with a repeated eigenvalue, which is either proportional to the identity or a nontrivial Jordan block. The former requires the couplings between the two colliding branches to vanish simultaneously with their diagonal mismatch, which does not occur generically and needs an additional constraint, such as a unitary symmetry under which the two branches belong to different sectors.

In particular, an isolated generic in-gap exceptional point cannot be removed by a small symmetry-preserving perturbation: the perturbation may move it, but the index must still change across it, so it can only disappear by merging with another domain wall or by reaching a band edge, where $\nu$ is not defined.

We note in passing that a collision between a decaying and a growing multiplier would require $|z|=1$ and therefore cannot occur inside the gap. Combined with the symmetry $z\mapsto 1/z$ of the transfer-matrix spectrum, this means that each in-gap exceptional energy carries two exceptional points of $T(E)$, one in $\mathcal{Z}_{<}$ and its mirror partner in $\mathcal{Z}_{>}$. Throughout we count exceptional energies rather than exceptional points of $T(E)$.

\subsubsection{Two-branch form}

When exactly two decaying branches participate, as in the four-band models of Sec.~\ref{sec: in-gap transition}, Eq.~\eqref{appendix: def D(E)} takes a transparent form in terms of momenta. Writing the two branches as $\lambda_\pm(E)=e^{\mathrm{i}k_\pm(E)}$ and
\begin{equation}
    \delta k(E):=k_+(E)-k_-(E),
\end{equation}
one finds
\begin{equation}
\label{appendix: two branch D}
    \mathcal{D}(E)=\left(\lambda_+-\lambda_-\right)^{2}
    =-4\,\lambda_+\lambda_-\sin^{2}\!\left(\frac{\delta k(E)}{2}\right),
\end{equation}
with $\lambda_+\lambda_-=e^{\mathrm{i}(k_++k_-)}$. Note that $\mathcal{D}$ vanishes if and only if $\lambda_+=\lambda_-$, that is only at a genuine degeneracy of the transfer matrix.

If the two momenta lie on the imaginary axis,
\begin{equation}
    k_{\pm}(E)=\mathrm{i}q_{\pm}(E),\qquad q_{\pm}(E)\in\mathbb{R},
\end{equation}
then $\lambda_{\pm}=e^{-q_{\pm}}$ are real and positive, $\delta k$ is purely imaginary, $\sin^{2}(\delta k/2)=-\sinh^{2}[(q_+-q_-)/2]<0$, and hence $\mathcal{D}>0$ and $\nu=0$. If instead the two momenta acquire opposite real parts,
\begin{equation}
    k_{\pm}(E)=\pm s(E)+\mathrm{i}q(E),\qquad s,q\in\mathbb{R},
\end{equation}
then $\lambda_{\pm}=e^{-q}e^{\pm\mathrm{i}s}$ form a complex-conjugate pair, $\lambda_+\lambda_-=e^{-2q}>0$, $\delta k=2s$ is real, and hence $\mathcal{D}<0$ and $\nu=1$. The index therefore changes across an in-gap exceptional point, as it must.

The same expression also characterises ordinary band-edge exceptional points, even though $\nu$ itself is not defined there because the colliding multipliers sit on the unit circle. Writing $k_{\pm}=k_0\pm s$ with $k_0=0$ or $k_0=\pi$, so that $\lambda_+\lambda_-=e^{2\mathrm{i}k_0}=1$, one finds $\mathcal{D}<0$ on the propagating side, where $s$ is real, and $\mathcal{D}>0$ on the evanescent side, where $s=\mathrm{i}\kappa$ is purely imaginary.

This construction is the transfer-matrix analogue of the discriminant $\mathbb{Z}_2$ characterisation of antiunitary-symmetry-protected exceptional points in non-Hermitian systems~\cite{Kawabata2019PRX,Kawabata2019PRL}.

\subsubsection{Relation to the point-gap invariant}

The point-gap invariant of Eq.~\eqref{eq: Z2 index} in the main text is evaluated with a reference point at a given EP, whereas $\nu(E)$ in Eq.~\eqref{appendix: def nu} is defined globally on the gap. The two agree near each in-gap EP. Let $z_{\mathrm{EP}}=e^{\mathrm{i}k_{\mathrm{EP}}}$ be the real multiplier at which two decaying multipliers $z_a$ and $z_b$ collide. Close to the EP, only this pair and its growing partners $1/z_a$, $1/z_b$ change character, so all other factors of $\det[z_{\mathrm{EP}}\mathbb{1}-T(E)]$ and of $\mathcal{D}(E)$ keep their signs. Where $z_a$ and $z_b$ are real, they lie on either side of $z_{\mathrm{EP}}$, so that
\begin{equation}
    (z_{\mathrm{EP}}-z_a)(z_{\mathrm{EP}}-z_b)<0,\qquad (z_a-z_b)^2>0,
\end{equation}
whereas where they form a complex-conjugate pair, $z_b=z_a^*$,
\begin{equation}
    (z_{\mathrm{EP}}-z_a)(z_{\mathrm{EP}}-z_b)=|z_{\mathrm{EP}}-z_a|^2>0,\qquad (z_a-z_b)^2<0.
\end{equation}
The growing partners contribute a positive factor to the determinant on both sides: when real, $1/z_a$ and $1/z_b$ have the same sign as $z_{\mathrm{EP}}$ but larger modulus, and otherwise they form a complex-conjugate pair. Hence, near the EP,
\begin{equation}
    \mathrm{sgn}\det\!\left[z_{\mathrm{EP}}\mathbb{1}-T(E)\right]=-C\,\mathrm{sgn}\,\mathcal{D}(E),
\end{equation}
where the sign $C$ is fixed by the remaining multipliers and does not change across the EP. With exactly two decaying branches, as in all models of this paper that host in-gap EPs, there are no remaining multipliers, $C=1$, and near each EP Eq.~\eqref{eq: Z2 index} coincides with Eq.~\eqref{appendix: def nu}. Away from the EP, the point gap at $z_{\mathrm{EP}}$ can also close when a single real multiplier crosses it, and different EPs generally have different $z_{\mathrm{EP}}$; the global labelling of the gap is therefore given by $\mathcal{D}(E)$.

\subsubsection{Gap phases and pair creation}

Because $\nu$ changes by one at each generic in-gap exceptional point, the number of energies at which we find EPs in a given gap has the same parity as $\nu(E_{-})+\nu(E_{+})$, where $E_{\mp}$ denote energies immediately inside the lower and upper edges of the gap. In particular, as long as $\nu(E_{-})$ and $\nu(E_{+})$ do not change, generic in-gap exceptional points are created and annihilated in pairs. The values $\nu(E_{\mp})$ can change only when an exceptional point passes through the corresponding band edge, which is how a single generic in-gap exceptional point can enter or leave the gap. Conversely, a gap in which $\nu$ differed at its two edges would necessarily contain an odd number of in-gap energies at which EPs are located.

The models considered in this paper all satisfy $\nu(E_-)=\nu(E_+)=0$ as long as no in-gap EPs are moved to the band edge. In the four-band Dirac model this happens at $b^4=c^2(a^2+b^2)$; beyond it, the band edges move to nonzero real momentum $\pm k_{\mathrm{min}}$, the two decaying multipliers form a complex-conjugate pair at every energy inside the gap, with $\mathrm{Re}\,k\to\pm k_{\mathrm{min}}$ at the band edges, and $\nu=1$ throughout the gap. At a band edge at $k_0\in\{0,\pi\}$ the two colliding branches leave the unit circle along the real axis, so the newly evanescent decaying multiplier is real and contributes no complex-conjugate pair; in the four-band models of Sec.~\ref{sec: in-gap transition} the remaining decaying multiplier is real immediately inside either edge as well. Explicitly, for $|b|<|c|$ one finds $\nu(E)=0$ throughout the gap, whereas for $|b|>|c|$ and $a\neq0$,
\begin{equation}
    \nu(E)=
    \begin{cases}
        1, & |E|<E_{\mathrm{EP}}^{+},\\[2pt]
        0, & E_{\mathrm{EP}}^{+}<|E|<E_{\mathrm{edge}},
    \end{cases}
\end{equation}
where $E_{\mathrm{edge}}=\big|\sqrt{a^2+b^2}-|c|\big|$ is the edge of the central gap, so that the region with $\nu=1$ is precisely the energy window bounded by the two in-gap exceptional points.

The two gap phases are therefore distinguished by whether the function $E\mapsto\nu(E)$ vanishes identically on the gap or is nonzero on a finite window, and this distinction is robust because $\nu$ can only change where the discriminant changes sign, i.e., where two real decaying multipliers turn into a complex-conjugate pair, which by the argument of Appendix~\ref{appendix: Existence of EP} is an exceptional point. It should nevertheless be understood in the same sense as the Weyl-semimetal analogy used in the main text: the $\nu=1$ window can be shrunk to zero width by annihilating its two bounding exceptional points against each other, without ever closing the propagating gap, or by pushing them into the band edges. What is protected is each individual exceptional point, as a domain wall of $\nu$ inside the gap, rather than the existence of the window itself.

\section{Properties of the Bloch Hamiltonian for complex $k$}
\label{appendix: props H_B in gap}
The Bloch Hamiltonian is given by 
\begin{equation}
    H_{\mathrm{B}}(k)=\sum_{l=0}^R t_le^{\mathrm{i}kl}+t_l^{\dagger}e^{-\mathrm{i}kl}.
\end{equation}
For real $k$, this Hamiltonian is Hermitian $H_{\mathrm{B}}(k)^\dagger=H_{\mathrm{B}}(k)$ and if spinless TRS is present then $H_{\mathrm{B}}(k)^*=H_{\mathrm{B}}(-k)$. However, if $k$ is not real then these two conditions are no longer true but they imply
\begin{equation}
\label{eq: condition 1 for non-H}
    \begin{aligned}
         H_{\mathrm{B}}(k)^\dagger&=\sum_{l=0}^R t^\dagger_l e^{-\mathrm{i}k^*l}+t_le^{\mathrm{i}k^*l}\\
         &=H_{\mathrm{B}}(k^*),
    \end{aligned}
\end{equation}
and
\begin{equation}
\label{eq: condition 2 for non-H}
    \begin{aligned}
        H_{\mathrm{B}}(k)^*&=\sum_{l=0}^R t^*_l e^{-\mathrm{i}k^*l}+(t^\dagger_l)^*e^{\mathrm{i}k^*l}\\
        &=H_{\mathrm{B}}(-k^*).
    \end{aligned}
\end{equation}
where we used that spinless TRS implies that the matrices $t_l$ are real. Therefore, what can we conclude for the spectrum of $H_{\mathrm{B}}(k)$ inside the gap? From the first condition
\begin{equation}
    \begin{aligned}
        H_{\mathrm{B}}(k)\phi(k)&=E(k)\phi(k)\\
        \phi^\dagger(k) H_{\mathrm{B}}(k)^\dagger&=E(k)\phi^\dagger(k)\\
        \phi^\dagger(k) H_{\mathrm{B}}(k^*)&=E(k)\phi^\dagger(k),
    \end{aligned}
\end{equation}
where we used that $E(k)\in\mathbb{R}$ for the complex $k$ that we are interested in. Since a left eigenvalue is also an eigenvalue, this shows that $E(k)$ belongs to the spectrum of $H_{\mathrm{B}}(k^*)$ as well. For the second condition
\begin{equation}
    \begin{aligned}
        H_{\mathrm{B}}(k)\phi(k)&=E(k)\phi(k)\\
        H_{\mathrm{B}}(k)^*\phi^*(k)&= E(k)\phi^*(k)\\
        H_{\mathrm{B}}(-k^*)\phi^*(k)&= E(k)\phi^*(k),
    \end{aligned}
\end{equation}
which shows that $E(k)$ is also an eigenvalue of $H_{\mathrm{B}}(-k^*)$. Taken together, the two conditions imply that if $E(k)$ is an eigenvalue of $H_{\mathrm{B}}(k)$, then it is an eigenvalue of $H_{\mathrm{B}}(k^*)$ and of $H_{\mathrm{B}}(-k^*)$ as well. Therefore, both $k^*$ and $-k^*$ solve Eq.~\eqref{eq: det equation for H}.

\section{Asymptotic behaviour of the bound-state tails from the transfer matrix}
\label{appendix: Asymptotic tails}
Here we show the relation between the eigensystem of the transfer matrix and the bulk behaviour of a bound state created by a local potential.
\subsection{Spatial profile of the bound state}
First, we consider a translationally invariant 1D lattice with hopping range $R$, $q$ orbitals and open boundary conditions. In the supercell basis this can be written as a nearest-neighbour hopping model
\begin{equation}
H_0=\sum_{r=1}^{N}M\ket{r}\bra{r}+\sum_{r=1}^{N-1}J\ket{r}\bra{r+1}+J^{\dagger}\ket{r+1}\bra{r},
\end{equation}
with $M$ and $J$ being $\mathcal{N}\times \mathcal{N}$ matrices, where $\mathcal{N}$ is the size of the chosen supercell, taken large enough that $J^2=0$ (see Appendix~\ref{appendix: construction general transfer matrix}). We now add a local onsite perturbation at the centre of the lattice,
\begin{equation}
    H_0\rightarrow H=H_0+U_0\ket{N/2}\bra{N/2},
\end{equation}
and assume that for some $U_0$, which is a square matrix of dimension $\mathcal{N}$, the perturbation produces a normalisable bound state with energy $E_{\mathrm{b}}$ in the bulk gap. Now, we write that energy eigenstate as 
\begin{equation}
    \ket{\Psi}=\sum_{r=1}^{N}\Psi_{r}\ket{r},
\end{equation}
with $\Psi_{r}\in\mathbb{C}^{\mathcal{N}}$. Then, its component at the site $N/2$ is given by the time-independent Schrödinger equation
\begin{equation}
    M\Psi_{N/2}+J\Psi_{N/2+1}+J^{\dagger}\Psi_{N/2-1}+U_0\Psi_{N/2}=E_{\mathrm{b}}\Psi_{N/2}.
\end{equation}
In the region that is far from both the impurity and the physical boundaries, the local potential is absent and the lattice is translationally invariant. Hence the wavefunction satisfies the homogeneous bulk equation
\begin{equation}
\label{eq: SE no impurity}
    M\Psi_{r}+J\Psi_{r+1}+J^{\dagger}\Psi_{r-1}=E_{\mathrm{b}}\Psi_{r}.
\end{equation}
The constant-coefficient recurrence admits Bloch-type solutions
\begin{equation}
    \Psi_{r}=z^{r}v,
\end{equation}
with $z\in\mathbb{C}$ and $v\in\mathbb{C}^{\mathcal{N}}$. Substitution into the homogeneous bulk equation gives
\begin{equation}
\label{eq: eq for z and v}
    \left( M+Jz+J^{\dagger}z^{-1}\right)v=E_{\mathrm{b}}v.
\end{equation}
Thus, $z$ and $v$ are determined entirely by the translationally invariant bulk problem.
Let $\Psi^{(j)}_r=z_j^r v_j$ denote linearly independent solutions of the homogeneous bulk equation at the same energy $E_{\mathrm{b}}$, i.e.
\begin{equation}
    h(z_j)v_j=E_{\mathrm{b}} v_j,
\end{equation}
with $h(z):=M+Jz+J^{\dagger}z^{-1}$. Since the Schrödinger equation is linear, any linear combination of such bulk modes at the same energy
\begin{equation}
\tilde{\Psi}_{r}:=\sum_{j}c_j\Psi^{(j)}_{r},
\end{equation}
 where the sum runs over all linearly independent solutions to the bulk equation, is again a solution. Thus, the solution for the bulk components of the spatial profile of the bound state (which is a normalisable eigenstate) can be written as
\begin{equation}
\label{app: eq expansion of eigenmodes}
    \Psi_r =
\begin{cases}
\displaystyle
\sum_{\alpha\in\mathcal D_R}
c_\alpha z_\alpha^{\,r-N/2}v_\alpha, & r>N/2,\\
\displaystyle
\sum_{\beta\in\mathcal D_L}
d_\beta z_\beta^{\,r-N/2}v_\beta, & r<N/2,
\end{cases}
\end{equation}
where $\mathcal D_R=\{\alpha: |z_\alpha|<1\}$ and
$\mathcal D_L=\{\beta: |z_\beta|>1\}$. The central amplitude $\Psi_{N/2}$ and the coefficients $c_\alpha,d_\beta$ are fixed by matching through the impurity region. They are therefore nonuniversal, whereas the possible decay lengths and oscillation wavevectors are determined by the bulk transfer-matrix spectrum at $E=E_{\mathrm{b}}$.

In an infinite lattice, Eq.~\eqref{app: eq expansion of eigenmodes} is exact: a mode growing away from the impurity with nonzero amplitude would make the state non-normalisable. In a finite lattice, the boundary conditions also require such modes, but with exponentially small amplitudes. For a single-band chain with nearest-neighbour hopping, for instance, the condition $\psi_{N+1} = 0$ gives the exact right tail $\psi_r = c z^{r - r_0}\left[1 - z^{2(N + 1 - r)}\right]$, with $|z| = e^{-q_i} < 1$, so the growing mode contributes at relative order $e^{-2q_i(N + 1 - r)}$. Moving a distance $d$ back into the bulk, the slowest-growing mode decreases by a factor $e^{-q_i d}$, since Hermiticity pairs it with the slowest-decaying mode as $z \leftrightarrow 1/z^*$, while the leading decaying mode increases by a factor $e^{q_i d}$. Their relative contribution is therefore of order $e^{-2q_i d}$, so that these modes are negligible in the bulk, the region shown in Fig.\ref{fig: Bound state}.

\subsection{Relation between the eigenvalues of the transfer matrix and the spatial multipliers $z_j$}
\label{appendix: spatial multipliers}
The following discussion applies to any energy $E$ for which the bulk transfer matrix is well defined. In the bulk, the reduced transfer matrix propagates the SVD components between neighbouring supercells according to
\begin{equation}
    T_{n,n+1}(E)
    \begin{pmatrix}
        \beta_n\\
        \alpha_{n-1}
    \end{pmatrix}
    =
    \begin{pmatrix}
        \beta_{n+1}\\
        \alpha_n
    \end{pmatrix}.
\end{equation}
The subscript $n,n+1$ indicates which components are propagated. In a translationally invariant bulk region, the transfer matrix is independent of $n$, and therefore
\begin{equation}
    T_{n,n+1}(E)=T(E).
\end{equation}

Using the SVD decomposition, the physical wavefunction in supercell $n$ is written as
\begin{equation}
    \Psi_n=U\alpha_n+W\beta_n+X\zeta_n.
\end{equation}
The component $\zeta_n$ is not independently propagated. Instead, it is fixed algebraically by $\alpha_{n-1}$ and $\beta_{n}$. Explicitly,
\begin{equation}
    \zeta_n=
    G_{UX}G_{UW}^{-1}\beta_n
    +
    \left(
    G_{WX}
    -
   G_{UX}G_{UW}^{-1}G_{WW}
    \right)\Sigma\alpha_{n-1}.
\end{equation}
where 
\begin{equation}
    G_{AB}:=B^{\dagger}(E-M)^{-1}A
\end{equation} for $A,B\in\{U,W,X\}$. Now consider the bulk transfer matrix eigensystem
\begin{equation}
    \begin{pmatrix}
        \beta_{n+1}\\
        \alpha_n
    \end{pmatrix}
    =
    T(E)
    \begin{pmatrix}
        \beta_n\\
        \alpha_{n-1}
    \end{pmatrix}
    =
    \lambda(E)
    \begin{pmatrix}
        \beta_n\\
        \alpha_{n-1}
    \end{pmatrix}.
\end{equation}
Therefore
\begin{equation}
    \beta_{n+1}=\lambda(E)\beta_n,
    \qquad
    \alpha_n=\lambda(E)\alpha_{n-1},
\end{equation}
for all $n$ in the bulk. Since $\zeta_n$ is a linear function of the propagated components, the auxiliary component scales in the same way:
\begin{equation}
    \zeta_{n+1}=\lambda(E)\zeta_n.
\end{equation}
Consequently, the physical bulk solution reconstructed from a single transfer eigenmode satisfies
\begin{equation}
    \Psi_{n+1}=\lambda(E)\Psi_n.
\end{equation}
It follows that
\begin{equation}
    \Psi_n=\lambda(E)^n \varphi
\end{equation}
for some internal vector $\varphi$.

Since this reconstructed $\Psi_n$ satisfies the homogeneous bulk Schrödinger equation, substituting $\Psi_n=\lambda(E)^n\varphi$ into
\begin{equation}
    M\Psi_n+J\Psi_{n+1}+J^\dagger\Psi_{n-1}=E\Psi_n
\end{equation}
gives
\begin{equation}
    \left(M+J\lambda(E)+J^\dagger\lambda(E)^{-1}\right)\varphi=E\varphi.
\end{equation}
Thus, at fixed energy $E$, every eigenvalue $\lambda(E)$ of the bulk transfer matrix is one of the spatial multipliers $z_j$ appearing in the exponential solutions of the homogeneous bulk equation. In the bound-state problem, this identification is evaluated at $E=E_{\mathrm{b}}$.

\subsection{Oscillatory tails from complex transfer-matrix momenta}

We now apply the previous result to the case where the bound-state energy $E_{\mathrm{b}}$ lies in the region between two in-gap exceptional points. The tails of the bound state are linear combinations of all normalisable vectors $\Psi^{j}_n$ constructed from the transfer-matrix modes at $E_{\mathrm{b}}$. However, sufficiently far from the impurity, the asymptotic behaviour is controlled by the modes with the slowest exponential decay.

Denote $r_0:=N/2$. We have shown that the right tail of the bound state can be expanded as 
\begin{equation}
    \Psi_r^{\mathrm{R}}=\sum_{\alpha\in\mathcal D_R}
    c_\alpha z_\alpha^{r-r_0}v_\alpha,
    \qquad |z_\alpha|<1.
\end{equation}
Since $|z_\alpha|<1$, far from the impurity, $\Psi_r^{\mathrm{R}}$ is dominated by those $z_\alpha$ with maximal absolute value. In other words, the bulk behaviour of $\Psi_r^{\mathrm{R}}$ is given by those $z_{\alpha}$ with smallest decay rate. Analogously, the asymptotic behaviour of the left tail is also dominated by the $z_{\beta}$ with the smallest decay rate.

 Let us call $k$ the associated momentum of the dominant $z$ of the right tail,
\begin{equation}
    k=q_r+\mathrm{i}q_i,
    \qquad q_r,q_i\in\mathbb{R},
    \qquad q_i>0,
\end{equation}
so that
\begin{equation}
    z=e^{\mathrm{i}k}=e^{-q_i}e^{\mathrm{i}q_r}
\end{equation}
is a right-decaying spatial multiplier, $|z|<1$.

For a spinless TRS system, the bulk Hamiltonian can be chosen real. Equivalently, the transfer matrix in the bulk may be chosen real. Therefore, if $z=e^{\mathrm{i}k}$ is an eigenvalue of the transfer matrix, then its complex conjugate $z^*=e^{-q_i}e^{-\mathrm{i}q_r}$ is also an eigenvalue. In terms of momenta, this corresponds to the pair
\begin{equation}
    k=q_r+\mathrm{i}q_i,
    \qquad
    -k^*=-q_r+\mathrm{i}q_i.
\end{equation}
Let $v$ be the eigenvector of the mode given by $z$, i.e.
\begin{equation}
    h(z)v=E_{\mathrm{b}}v.
\end{equation}
Then, due to TRS, $v^*$ is an eigenvector of $h(z^*)$ with eigenvalue $E_{\mathrm{b}}$. From our previous analysis, we know that the leading part of the right tail has therefore the form
\begin{equation}
    \Psi_r^{{\mathrm{R}}}
    =
    c_1z^{r-r_0}v
    +
    c_2\,(z^*)^{r-r_0}v^*,
    \qquad r>r_0,
\end{equation}

Since the Hamiltonian is real, a nondegenerate bound state may be chosen real. Imposing $\Psi_r^{\mathrm{R}}=(\Psi_r^{\mathrm{R}})^*$ gives
\begin{equation}
     c_2=c_1^*.
\end{equation}
Therefore the right tail can be written as
\begin{equation}
    \Psi_r^{{\mathrm{R}}}
    =
    2\mathrm{Re}\left[
    c_1z^{r-r_0}v
    \right].
\end{equation}
Writing $c_1=|c_1|e^{\mathrm{i}\phi}$ and $v_\mu=|v_\mu|e^{\mathrm{i}\theta_\mu}$ for the $\mu$-th orbital component, we obtain
\begin{equation}
    \Psi_{r,\mu}^{{\mathrm{R}}}
    =
    2|c_1||v_\mu|
    e^{-q_i(r-r_0)}
    \cos\left[
    q_r(r-r_0)+\phi+\theta_\mu
    \right],
\end{equation}
for $r>r_0$. Thus the right tail is not merely exponentially decaying: it is an exponentially damped oscillation. The decay length is fixed by $q_i=\mathrm{Im}\,k$, while the oscillation wavevector is fixed by $q_r=\mathrm{Re}\,k$. The overall amplitude and phase are fixed by matching through the impurity region.

The same reasoning applies to the left tail. With the convention $z^{r-r_0}$, decay as $r\to -\infty$ requires $|z|>1$. The left-decaying partners associated with the same complex momentum may be written as
\begin{equation}
    z_{\mathrm{L}}=e^{-\mathrm{i}k}=e^{q_i}e^{-\mathrm{i}q_r},
    \qquad
    z_{\mathrm{L}}^*=e^{\mathrm{i}k^*}=e^{q_i}e^{\mathrm{i}q_r}.
\end{equation}
Although $|z_{\mathrm{L}}|>1$, the factor $z_{\mathrm{L}}^{r-r_0}$ decays for $r<r_0$, because $r-r_0<0$. Therefore the leading left tail has the form
\begin{equation}
    \Psi_r^{{\mathrm{L}}}
    =
    d\,z_{\mathrm{L}}^{r-r_0}w
    +
    d^*(z_{\mathrm{L}}^*)^{r-r_0}w^*,
    \qquad r<r_0,
\end{equation}
or equivalently
\begin{equation}
\label{app: eq left tail}
    \Psi_r^{{\mathrm{L}}}
    =
    2\mathrm{Re}\left[
    d\,z_{\mathrm{L}}^{r-r_0}w
    \right],
\end{equation}
where $w$ is the eigenvector associated with the eigenvalue $z_{\mathrm{L}}$. Writing $d=|d|e^{\mathrm{i}\varphi}$ and $w_\mu=|w_\mu|e^{\mathrm{i}\vartheta_\mu}$ gives
\begin{equation}
    \Psi_{r,\mu}^{{\mathrm{L}}}
    =
    2|d||w_\mu|
    e^{-q_i(r_0-r)}
    \cos\left[
    q_r(r_0-r)+\varphi+\vartheta_\mu
    \right],
\end{equation}
for $r<r_0$. Hence both tails oscillate whenever the leading decaying transfer-matrix mode present in the bound-state profile has $\mathrm{Re}\,k\neq0,\pm\pi$.

By contrast, when the relevant decaying transfer-matrix eigenvalues are real, the corresponding momenta have $\mathrm{Re}\,k=0$ or $\pm\pi$. In that case the leading tail is purely exponential. Therefore the appearance of a nonzero real part of the in-gap momentum between exceptional points gives a direct physical signature in the spatial profile of impurity bound states.

\subsection{Tails at an in-gap exceptional point}

The expansion of Eq.~\eqref{app: eq expansion of eigenmodes} assumes that the bulk solutions at $E_{\mathrm{b}}$ are spanned by modes of the form $z_j^{r} v_j$. This fails at an in-gap exceptional energy. There two decaying multipliers coalesce at a real value $z_{\mathrm{EP}} = e^{\mathrm{i}k_{\mathrm{EP}}}$, and $T(E_{\mathrm{b}})$ is defective (Appendix~\ref{app: defectiveness of T}): it has a single eigenvector $x$ for this eigenvalue, which gives the bulk solution $\Psi_n = z_{\mathrm{EP}}^{n} v$ with $h(z_{\mathrm{EP}})v = E_{\mathrm{b}} v$. The second solution associated with the coalescing pair follows from a generalised eigenvector $y$,
\begin{equation}
  T(E_{\mathrm{b}}) y = z_{\mathrm{EP}} y + x .
\end{equation}
Propagating $y$ with the transfer matrix, by induction
\begin{equation}
  T(E_{\mathrm{b}})^n y = z_{\mathrm{EP}}^{n} y + n z_{\mathrm{EP}}^{n-1}\, x .
  \label{eq:Td}
\end{equation}
Since $\Psi_n$ is fixed linearly by the propagated components $(\beta_n, \alpha_{n-1})$, and $x$ reconstructs to $v$, the corresponding solution of the bulk Eq.~\eqref{eq: SE no impurity} is
\begin{equation}
  \Psi_n = z_{\mathrm{EP}}^{n} \tilde v + n z_{\mathrm{EP}}^{n-1} v ,
\end{equation}
where $\tilde v$ is the vector reconstructed from $y$.

At $E_{\mathrm{b}} = E_{\mathrm{EP}}$, the right tail therefore becomes
\begin{equation}
\begin{aligned}
  \Psi^{\mathrm{R}}_r = &c_1 z_{\mathrm{EP}}^{r-r_0} v
   + c_2 \Big[ z_{\mathrm{EP}}^{r-r_0}\, \tilde v
   + (r-r_0) z_{\mathrm{EP}}^{r-r_0-1}\, v \Big]\\
   &+ \sum_{\alpha} c_\alpha z_\alpha^{r-r_0} v_\alpha ,
   \end{aligned}
\end{equation}
for $r > r_0$, where the sum runs over the remaining decaying modes, and all coefficients are fixed by matching through the impurity region. If the coalescing pair is the slowest-decaying one, $|z_\alpha| < |z_{\mathrm{EP}}|$, and $c_2 \neq 0$, then far from the impurity the term proportional to $r-r_0$ dominates, and
\begin{equation}
  \Psi^{\mathrm{R}}_r \propto (r-r_0) e^{\mathrm{i}k_{\mathrm{EP}}(r-r_0)} v .
\end{equation}
The tail thus decays neither purely exponentially nor as a damped standing wave.

The left tail follows in the same way from the mirror exceptional point of $T(E_{\mathrm{b}})$ at $z_{\mathrm{L}} = 1/z_{\mathrm{EP}}$ in the growing subspace: $\Psi^{\mathrm{L}}_r \propto (r_0-r) e^{\mathrm{i}k_{\mathrm{EP}}(r_0-r)} w$, with $w$ the eigenvector associated with $z_{\mathrm{L}}$ as in Eq.~\eqref{app: eq left tail}.

\section{Transfer matrix for a photonic crystal}
\label{appendix: construction of T for PC}

Here we study the case in which the dielectric constant is replaced by a matrix called the dielectric tensor
\begin{equation}
    \epsilon\rightarrow \begin{pmatrix}
        \epsilon_{xx} & \epsilon_{xy} & \epsilon_{xz}\\
        \epsilon_{yx} & \epsilon_{yy} & \epsilon_{yz}\\
        \epsilon_{zx} & \epsilon_{zy} & \epsilon_{zz}
    \end{pmatrix}.
\end{equation}
The reference frame in which the dielectric tensor is diagonal is defined by the material's so-called principal axes. From Faraday's law and Ampère-Maxwell law, we obtain
\begin{equation}
\label{eq: Faraday's law}
    \begin{aligned}
        \nabla\times E &=-\frac{\partial B}{\partial t}\\
        \nabla\times(\nabla\times E)&=-\frac{\partial^2 D}{\partial t^2},
    \end{aligned}
\end{equation}
where we set $c=\mu_0=\epsilon_0=1$. We consider an incident monochromatic displacement field $D$ propagating along the $z$-direction with frequency $\omega$
\begin{equation}
\label{appendix: phasors}
    D(z,t)=D_{\omega}(z)e^{-\mathrm{i}\omega t}+\mathrm{c.c.}
\end{equation}
Then, the continuity conditions correspond to
\begin{equation}
    \begin{aligned}
E_{\mathrm{out},\parallel}(x,y,0,t)&=E_{\mathrm{in},\parallel}(x,y,0,t),\\
 H_{\mathrm{out},\parallel}(x,y,0,t)&=H_{\mathrm{in},\parallel}(x,y,0,t),\\
 D_{\mathrm{out},z}(x,y,0,t)&=D_{\mathrm{in},z}(x,y,0,t),
 \end{aligned}
\end{equation}
for all $x,y$ and $t$ (there are no free charges or surface currents). This implies that for plane wave solutions inside the material, we get
\begin{equation}
    \begin{aligned}
        k^{\mathrm{in}}_x&=k^{\mathrm{out}}_x,\\
        k^{\mathrm{in}}_y&=k^{\mathrm{out}}_y,\\
        \implies k^{\mathrm{in}}_x&=k^{\mathrm{in}}_y=0,
    \end{aligned}
\end{equation}
since outside the material the wavevector is just $(0,0,k^{\mathrm{out}})$ and that the frequency of the inside waves is also $\omega$. Thus, inside the material the curl--curl equation, Eq.~\eqref{eq: Faraday's law}, corresponds to
\begin{equation}
    k^2_{\mathrm{in}}\begin{pmatrix}
        E_{x,0}\\E_{y,0}\\0
    \end{pmatrix}=\omega^2\epsilon \begin{pmatrix}
        E_{x,0}\\E_{y,0}\\E_{z,0}
    \end{pmatrix},
\end{equation}
where we have defined a single mode given by
\begin{equation}
    D_{\omega,\mathrm{in}}(z)=\epsilon_{\mathrm{in}}\begin{pmatrix}
         E_{x,0}\\E_{y,0}\\ E_{z,0}
    \end{pmatrix} e^{\mathrm{i}k_{\mathrm{in}}z},
\end{equation}
such that we work only with one complex phasor $D_{\omega}(z)$. From now on, we omit the subscript $\mathrm{in}$. Then, because there are no free charges, from Gauss's law we get that the $z$-component of the displacement field is 0:
\begin{equation}
\begin{aligned}
   \nabla \cdot D&=0\\
   kD_{z}&=0\\
\implies\sum_j \epsilon_{zj}E_j&=0.
\end{aligned}  
\end{equation}
Plugging this into Eq.~\eqref{eq: Faraday's law} we end up with the eigenvalue equation
\begin{equation}
    \frac{k^2}{\omega^2}\begin{pmatrix}
            E_{x,0}\\
        E_{y,0}
        \end{pmatrix}=\tilde{\epsilon}\begin{pmatrix}
            E_{x,0}\\
        E_{y,0}
        \end{pmatrix},
\end{equation}
where
\begin{equation}
    \tilde{\epsilon}:=\begin{pmatrix}
        \epsilon_{xx}-\epsilon_{xz}\epsilon_{zx}/\epsilon_{zz} & \epsilon_{xy}-\epsilon_{xz}\epsilon_{zy}/\epsilon_{zz}\\
        \epsilon_{yx}-\epsilon_{yz}\epsilon_{zx}/\epsilon_{zz} & \epsilon_{yy}-\epsilon_{yz}\epsilon_{zy}/\epsilon_{zz}
    \end{pmatrix},
\end{equation}
is a real Hermitian matrix as the material is considered lossless in any spatial layer and the system is considered to have TRS. Hence, given a certain $\omega$, the possible $k$ are obtained by computing the eigenvalues of $\tilde{\epsilon}$ which correspond to $k^2_i/\omega^2$. Then, the general solution for the displacement field corresponds to
\begin{equation}
\begin{aligned}
\label{eq: new electric field}
    D(z,t) &=
    \begin{pmatrix}
        \tilde{\epsilon}E_{2\mathrm{D}}(z,t)\\
        0
    \end{pmatrix}, \\
    E_{2\mathrm{D}}(z,t) &= e^{-\mathrm{i}\omega t}\Bigg[
    v_1
    \left(c_1e^{\mathrm{i}k_1 z}+c_2e^{-\mathrm{i}k_1 z}\right) \\
    &\qquad\qquad
    +
    v_2
    \left(d_1e^{\mathrm{i}k_2 z}+d_2e^{-\mathrm{i}k_2 z}\right)
    \Bigg].
\end{aligned}
\end{equation}
where $v_1$ and $v_2$ are the eigenvectors associated with $k_{1}$ and $k_{2}$ given by the different eigenvalues of $\tilde{\epsilon}$.
Given that solution, the continuity conditions can be written in matrix form. For a nonmagnetic medium with propagation along $z$, Faraday's law gives $H_{\parallel}\propto\hat{z}\times\partial_z E_{\parallel}$, so continuity of the in-plane magnetic field is equivalent to continuity of $\mathrm{d}E_{2\mathrm{D}}/\mathrm{d}z$; this is why the matching matrix $\mathcal{A}$ below contains both the field amplitudes and their $z$-derivatives. Matching at the interface between layers $j$ and $j+1$ then reads
\begin{equation}
    \mathcal{A}(\tilde{\epsilon}_j,\tilde{a}_j)\bs{\Psi}_j(\omega)=\mathcal{A}(\tilde{\epsilon}_{j+1},\tilde{a}_j)\bs{\Psi}_{j+1}(\omega),
\end{equation}
where
\begin{equation}
    \begin{aligned}
        \bs{\Psi}_j(\omega)&:=\begin{pmatrix}
            c_{1,j}\\c_{2,j}\\d_{1,j}\\d_{2,j}
        \end{pmatrix},\\
        \tilde{a}_j&:=\sum\limits_{i=1}^{j}a_i,\\
\mathcal A(\tilde\epsilon,z)&:=
\begin{pmatrix}
A_1 & A_2
\end{pmatrix},\\
A_\ell&:=
\begin{pmatrix}
E_{x,\ell}e^{\mathrm{i}k_\ell z} & E_{x,\ell}e^{-\mathrm{i}k_\ell z}\\
E_{y,\ell}e^{\mathrm{i}k_\ell z} & E_{y,\ell}e^{-\mathrm{i}k_\ell z}\\
k_\ell E_{x,\ell}e^{\mathrm{i}k_\ell z} & -k_\ell E_{x,\ell}e^{-\mathrm{i}k_\ell z}\\
k_\ell E_{y,\ell}e^{\mathrm{i}k_\ell z} & -k_\ell E_{y,\ell}e^{-\mathrm{i}k_\ell z}
\end{pmatrix},
\qquad \ell=1,2,
\end{aligned}
\end{equation}
 and the subscript $j$ indicates the coefficients of the $j$th layer. By this construction, the matrix given by the product 
 \begin{equation}
 [\mathcal{A}(\tilde{\epsilon}_{j+1},\tilde{a}_j)]^{-1}\mathcal{A}(\tilde{\epsilon}_j,\tilde{a}_j),
 \end{equation}
 is block diagonal if the eigenvectors of $\tilde{\epsilon}_j$ and $\tilde{\epsilon}_{j+1}$ are the same. Then, using Bloch's theorem we get that
\begin{equation}
\label{eq_app: FTeq}
\begin{pmatrix}
    E_{2\mathrm{D}}(z+a)\\\frac{\mathrm{d}E_{2\mathrm{D}}}{\mathrm{d}z}(z+a)
\end{pmatrix}=e^{\mathrm{i}k(\omega)a}\begin{pmatrix}
    E_{2\mathrm{D}}(z)\\ \frac{\mathrm{d}E_{2\mathrm{D}}}{\mathrm{d}z}(z)
\end{pmatrix},
\end{equation}
where $k(\omega)$ corresponds to crystal momentum. Consequently, we write 
\begin{equation}
\mathcal{A}(\tilde{\epsilon}_{n+1},\tilde{a}_n)\bs{\Psi}_{n+1}(\omega)=\mathcal{A}(\tilde{\epsilon}_1,0)\bs{\Psi}_1(\omega) e^{\mathrm{i}k(\omega)a}.
\end{equation}

Then, we get the following eigenvalue equation for the crystal momentum
\begin{equation}
    \begin{aligned}
       \mathcal{T}(\omega)\bs{\Psi}_1(\omega)&=e^{\mathrm{i}k(\omega)a}\bs{\Psi}_1(
       \omega
       ),\\
        \mathcal{T}(\omega)&:=\left[\mathcal{A}(\tilde{\epsilon}_1,0)\right]^{-1} \prod\limits_{j=n}^{2} \mathcal{A}(\tilde{\epsilon}_j,\tilde{a}_j)[\mathcal{A}(\tilde{\epsilon}_j,\tilde{a}_{j-1})]^{-1} \\
        &\times \mathcal{A}(\tilde{\epsilon}_1,a_1).
    \end{aligned}
\end{equation}

\section{Properties of the transfer matrix for a PhC from a general perspective}
\label{appendix: properties of Y for PC}
We start by considering the equation
\begin{equation}
\label{eq_app: eq for E_2D}
    \frac{\mathrm{d}^2E_{2\mathrm{D}}}{\mathrm{d} z^2}(z)=-\omega^2\tilde{\epsilon}E_{2\mathrm{D}}(z).
\end{equation}
Equivalently, we write
\begin{equation}
    \frac{\mathrm{d}^2E_{2\mathrm{D}}}{\mathrm{d}z^2}(z)=-M(\omega)E_{2\mathrm{D}}(z),
\end{equation}
where $M(\omega):=\omega^2\tilde{\epsilon}$. The general solution of this equation can be written as
\begin{equation}
    E_{2\mathrm{D}}(z)=\sum_{i=1}^2v_{k_i}\sum_{j=1}^2 c_j^{(i)}f_j^{(i)}(z),
\end{equation}
where $v_{k_i}$ is the eigenvector associated with the eigenvalue $k_i^2$, $f_j^{(i)}$ are the linearly independent solutions to the scalar equation
\begin{equation}
\label{eq_app: eq for f}
    \frac{\mathrm{d}^2f}{\mathrm{d}z^2}(z)=-k_i^2f(z),
\end{equation}
and $c_j^{(i)}$ are complex coefficients.
Now, we know that the transfer matrix for an anisotropic PhC propagates the vector constructed from the coefficients $c_j^{(i)}$ for the electric field $E_{2\mathrm{D}}$. However, since Eq.~\eqref{eq_app: eq for E_2D} has only real coefficients, if $E_{2\mathrm{D}}$ is a solution, then $E_{2\mathrm{D}}^*$ is also a solution and can be written as
\begin{equation}
\begin{aligned}
    E_{2\mathrm{D}}(z)^*=\sum_{i=1}^2v_{k_i}^*\sum_{j=1}^2 c_j^{*(i)}f_j^{*(i)}(z).
    \end{aligned}
\end{equation}
This property follows from the TRS of Maxwell's equations. If $E(r,t),D(r,t),B(r,t)$ and $H(r,t)$ solve Maxwell's equations, then 
\begin{equation}
\begin{aligned}
    e(r,t)&:=E(r,-t),\\
    d(r,t)&:=D(r,-t),\\
    b(r,t)&:=-B(r,-t),\\
    h(r,t)&:=-H(r,-t),
    \end{aligned}
\end{equation}
also solve Maxwell's equations. This implies that if we have monochromatic planes and we express our fields as phasors as in Eq.~\eqref{appendix: phasors}, then the equation satisfied by $D_{\omega}(z)$ is also satisfied by $D^*_{\omega}(z)$, provided that $\epsilon$ is real. More specifically, if $f$ is a solution of a differential equation with real coefficients, then $f^*$ also is.

Since $M(\omega)$ is a real Hermitian matrix, we can choose without loss of generality only real eigenvectors. Then, since Eq.~\eqref{eq_app: eq for f} has real coefficients, if $f$ is a solution, then $f^*$ is also a solution and thus, can be written as a linear combination of the known solutions, hence
\begin{equation}
\begin{aligned}
    E_{2\mathrm{D}}(z)^*&=\sum_{i=1}^2v_{k_i}\sum_{j=1}^2 c_j^{*(i)}\sum_{s=1}^{2}\sigma_{js}^{(i)}f_s^{(i)}(z)\\
    &=\sum_{i=1}^2v_{k_i}\sum_{s=1}^2 d_s^{(i)}f_s^{(i)}(z),
    \end{aligned}
\end{equation}
where we defined
\begin{equation}
    d_s^{(i)}:=\sum_{j=1}^2\left(\sigma^{(i)T}\right)_{sj}c_j^{*(i)}.
\end{equation}
Thus, we can construct the same equation for the vector $\tilde{\bs{\Psi}}$ given by the coefficients $d_s^{(i)}$ such that
\begin{equation}
\begin{aligned}
    \begin{pmatrix}
        d_1^{(1)}\\d_2^{(1)}\\d_1^{(2)}\\d_2^{(2)}
    \end{pmatrix}&=\begin{pmatrix}
        \left(\sigma^{(1)}\right)^{T}&0\\0&\left(\sigma^{(2)}\right)^{T}
    \end{pmatrix}\mathcal{K}\begin{pmatrix}
        c_1^{(1)}\\c_2^{(1)}\\c_1^{(2)}\\c_2^{(2)}
    \end{pmatrix},
    \end{aligned}
\end{equation}
which we write as $\tilde{\bs{\Psi}}=\tilde\sigma \mathcal{K} \bs{\Psi}$. By taking the complex conjugate $E_{2\mathrm{D}}(z)^*$, we use the same method of writing $f_s^{(i)*}(z)$ as linear combinations of $f_1^{(i)}(z)$ and $f_2^{(i)}(z)$, which gives
\begin{equation}
    \sum_{s=1}^2\sum_{j=1}^2c^{(i)}_j\sigma^{(i)*}_{js}\sigma^{(i)}_{sm}=c^{(i)}_m.
\end{equation}
This equation must be valid for any initial conditions and thus,
\begin{equation}
    \sigma^{(i)*}\sigma^{(i)}=\mathbb{1}.
\end{equation}
Consequently, we get that $(\tilde\sigma \mathcal{K})^2=\mathbb{1}$.
The vector $\tilde{\bs{\Psi}}$ is associated with a crystal momentum $-k^*$ which comes from complex conjugating Eq.~\eqref{eq_app: FTeq}. Thus, the eigenvalue equation of the transfer matrix for this vector becomes
\begin{equation}
    \mathcal{T}(\omega)\tilde{\bs{\Psi}}_1(\omega)=e^{-\mathrm{i}k^*(\omega)a}\tilde{\bs{\Psi}}_1(\omega),
\end{equation}
which implies that
\begin{equation}
\begin{aligned}
    \mathcal{T}(\omega) \tilde\sigma \mathcal{K} \bs{\Psi}_1(\omega)&=e^{-\mathrm{i}k^*(\omega)a}\tilde\sigma \mathcal{K} \bs{\Psi}_1(\omega)\\
    \tilde\sigma \mathcal{K}\mathcal{T}(\omega) \tilde\sigma \mathcal{K} \bs{\Psi}_1(\omega)&=e^{\mathrm{i}k(\omega)a}\bs{\Psi}_1(\omega)\\
    \implies \tilde\sigma \mathcal{K}\mathcal{T}(\omega) \tilde\sigma \mathcal{K}&=\mathcal{T}(\omega) ,
    \end{aligned}
\end{equation}
which corresponds to what we call antiunitary symmetry.
\subsection{Symplectic condition}
Now we see how the symplectic condition comes from a general argument. 
Consider the canonical variables $u(\lambda)$ and $v(\lambda)$ where both variables live in $\mathbb{R}^n$ for some natural $n$ for all $\lambda$ in some real interval. Here, $\lambda$ is a flow parameter that need not be the parameter of time. A Hamiltonian system is one in which there exists a so-called Hamiltonian function $H=H(u,v)$ such that
\begin{equation}
    \begin{aligned}
        u'&=\frac{\partial H}{\partial v},\\
        v'&=-\frac{\partial H}{\partial u},
    \end{aligned}
\end{equation}
where $'$ denotes $\mathrm{d}/\mathrm{d}\lambda$. This equation can be written more compactly as
\begin{equation}
    \eta'=\mathcal{J}\partial_\eta H,
\end{equation}
where 
\begin{equation}
    \mathcal{J}=\begin{pmatrix}
        0&\mathbb{1}\\-\mathbb{1} & 0
    \end{pmatrix}.
\end{equation}
 Now we say that there is a canonical transformation $\xi=\xi(\eta)$ if there is a new Hamiltonian-type function $\tilde H$ such that 
\begin{equation}
\label{eq: Ham syst for xi}
    \xi'=\mathcal{J}\partial_\xi \tilde H.
\end{equation}
There are standard criteria for determining if a transformation is canonical or not without actually finding Eq.~\eqref{eq: Ham syst for xi}. In short, a transformation is canonical if and only if the Poisson bracket is left invariant under the transformation, or equivalently, if the matrix $\mathcal{T}$ whose components are $\mathcal{T}_{ij}:=\frac{\partial \xi_i}{\partial\eta_j}$ satisfies the symplectic condition
\begin{equation}
    \mathcal{T}\mathcal{J}\mathcal{T}^{\mathrm{T}}=\mathcal{J}.
\end{equation}
Now, we show that the transformation given by the evolution of a Hamiltonian system, i.e. $\xi(\lambda):=\eta(\lambda+\Delta \lambda)$ is a canonical transformation. For that we compute the Poisson bracket for a transformation given by an infinitesimal translation in $\lambda$: $q(\lambda):=u(\lambda+\Delta \lambda)$ and $p(\lambda):=v(\lambda+\Delta \lambda)$, where we divided the interval from $\lambda$ to $\lambda+\Lambda$ (another point in the parameter space) into $N$ subintervals $\Delta\lambda$. Expanding around $\lambda$:
\begin{equation}
\begin{aligned}
    q(\lambda)&=u(\lambda)+u'(\lambda)\Delta \lambda\\
    &=u(\lambda)+\frac{\partial H}{\partial v}\Delta \lambda,\\
    p(\lambda)&=v(\lambda)+v'(\lambda)\Delta \lambda\\
    &=v(\lambda)-\frac{\partial H}{\partial u}\Delta \lambda.
\end{aligned}
\end{equation}
Thus,
\begin{equation}
    \begin{aligned}
        \{q_i,p_j\}_{u,v}&=\sum_{k}\left(\frac{\partial q_i}{\partial u_k}\frac{\partial p_j}{\partial v_k}-\frac{\partial q_i}{\partial v_k}\frac{\partial p_j}{\partial u_k}\right)\\
        &=\sum_{k}\bigg[\left(\delta_{ik}+\frac{\partial^2 H}{\partial u_k\partial v_i}\Delta \lambda\right)\left( \delta_{jk}-\frac{\partial^2 H}{\partial v_k\partial u_j}\Delta \lambda\right)\\
        &\qquad+\frac{\partial^2H}{\partial v_k\partial v_{i}}\frac{\partial^2H}{\partial u_k\partial u_j}(\Delta \lambda)^2\bigg]\\
        &=\delta_{ij}+\left\{\frac{\partial H}{\partial u_j},\frac{\partial H}{\partial v_i}\right\}_{u,v}(\Delta \lambda)^2\\
        &=\delta_{ij}+O\!\left((\Delta \lambda)^2\right).
    \end{aligned}
\end{equation}
In the same way, $\{q_i,q_j\}_{u,v}$ and $\{p_i,p_j\}_{u,v}$ vanish up to $O\!\left((\Delta\lambda)^2\right)$, since their first-order terms, $\Delta\lambda\,(\partial^2 H/\partial v_i\partial v_j-\partial^2 H/\partial v_j\partial v_i)$ and $\Delta\lambda\,(\partial^2 H/\partial u_i\partial u_j-\partial^2 H/\partial u_j\partial u_i)$, cancel.
Then, we can transform again to $Q(\lambda):=q(\lambda+\Delta \lambda)$ and $P(\lambda):=p(\lambda+\Delta \lambda)$ and so on, for $N$ steps.

Each of the $N=\Lambda/\Delta\lambda$ steps changes the brackets only at order $(\Delta\lambda)^2$, so the accumulated change is of order $\Lambda\,\Delta\lambda$. Therefore, in the limit $N\rightarrow \infty$, the Poisson brackets stay the same and consequently, the transformation is canonical.
Now, if the evolution is given by a matrix $\xi=\mathcal{T}\eta$, then $\mathcal{T}$ has to satisfy the symplectic condition, which implies that $\mathcal{T}^T$ is similar to the inverse of $\mathcal{T}$, i.e.
\begin{equation}
    \mathcal{T}^{\mathrm{T}}=\mathcal{J}^{-1}\mathcal{T}^{-1}\mathcal{J}.
\end{equation}
We also know that every square matrix is similar to its transpose. Hence, $\mathcal{T}\sim \mathcal{T}^{-1}$.
Coming back to the anisotropic PhC we can put Eq.~\eqref{eq_app: eq for E_2D} into a Hamiltonian form
\begin{equation}
    \frac{\mathrm{d}}{\mathrm{d}z}\begin{pmatrix}
    \bs{E}_{2\mathrm{D}}(z)\\ \frac{d\bs{E}_{2\mathrm{D}}}{dz}(z)
\end{pmatrix}=\begin{pmatrix}
    0&\mathbb{1}\\
    -M(\omega)&0
\end{pmatrix}\begin{pmatrix}
    \bs{E}_{2\mathrm{D}}(z)\\ \frac{d\bs{E}_{2\mathrm{D}}}{dz}(z)
\end{pmatrix},
\end{equation}
which, since its coefficients are real, has Hamiltonian form for the real and imaginary parts of $u=E_{2\mathrm{D}}$ and $v=\mathrm{d}E_{2\mathrm{D}}/\mathrm{d}z$ separately, with $H=\tfrac{1}{2}\left(v^{\mathrm{T}}v+u^{\mathrm{T}}M(\omega)u\right)$, provided that $M(\omega)$ is symmetric. This holds here, because $\tilde{\epsilon}$ is real and Hermitian, and hence symmetric, for the lossless media with TRS considered in Appendix~\ref{appendix: construction of T for PC}. The matrix that propagates $\left(E_{2\mathrm{D}},\mathrm{d}E_{2\mathrm{D}}/\mathrm{d}z\right)$ across one period therefore satisfies the symplectic condition and is similar to its inverse. Since $\mathcal{T}(\omega)$ is related to this matrix by a change of basis, which is invertible for $\omega\neq0$ and invertible $\tilde{\epsilon}$, so that $k_\ell\neq0$, also $\mathcal{T}(\omega)\sim\mathcal{T}(\omega)^{-1}$.

\end{document}